\documentclass[a4paper,onecolumn,11pt,accepted=2026-01-14]{quantumarticle}
\pdfoutput=1
\usepackage[utf8]{inputenc}
\usepackage[english]{babel}
\usepackage[T1]{fontenc}
\usepackage{amsmath, amsthm}
\usepackage{hyperref}

\usepackage{tikz}
\usepackage{lipsum}

\newtheorem{theorem}{Theorem}

\usepackage{amsfonts, amssymb}
\usepackage{xurl}

\usepackage{physics}
\usepackage{mathtools}
\usepackage{subcaption}
\usepackage{tikz}
\usetikzlibrary{quantikz}

\usepackage[numbers]{natbib}
\usepackage[resetlabels]{multibib}

\newcites{article}{article references}
\newcites{book}{book references}
\newcites{misc}{misc references}
\newcites{repo}{repository references}
\newcites{web}{website references}
\newcites{other}{Other references}

\newcommand{\logicX}{\mathcal{X}}
\newcommand{\logicZ}{\mathcal{Z}}

\newcommand{\F}{\mathbb{F}}
\newcommand{\CNOT}{\textsc{cnot}}

\begin{document}

\selectlanguage{english}

\title{A 3D lattice defect and efficient computations in topological MBQC}
\author[1,2]{Gabrielle Tournaire}
\author[3]{Marvin Schwiering}
\author[2,3]{Robert Raussendorf}
\author[1,4]{Sven Bachmann}

\affil[1]{{Department of Physics and Astronomy, University of British Columbia Vancouver, Canada}}

\affil[2]{{Stewart Blusson Quantum Matter Institute, University of British Columbia Vancouver, Canada}}

\affil[3]{Institut f{\"u}r Theoretische Physik, Leibniz Universit{\"a}t Hannover, Germany}

\affil[4]{Department of Mathematics, University of British Columbia Vancouver, Canada}

\begin{abstract}
    We describe an efficient, fully fault-tolerant implementation of Measurement-Based Quantum Computation (MBQC) in the 3D cluster state. The two key novelties are (i) the introduction of a lattice defect in the underlying cluster state and (ii) the use of the Rudolph-Grover rebit encoding. Concretely, (i) allows for a topological implementation of the Hadamard gate, while (ii) does the same for the phase gate. Furthermore, we develop general ideas towards circuit compaction and algorithmic circuit verification, which we implement for the Reed-Muller code used for magic state distillation. Our performance analysis highlights the overall improvements provided by the new methods. 
\end{abstract}

\section{Introduction}

Decoherence is an obstacle to scaling up quantum computers. It was once thought fundamental \cite{Raimond:1996pur}, but was later shown to be surmountable by the techniques of fault-tolerant quantum computation \cite{CalderankShorPeter_Good_ECC, Steane_Simple_QECC, Gottesman_thesis}. The central result in this regard is the Threshold Theorem \cite{AharonovBenOr_FTQCconstanterror, Aliferis2005QuantumAT}, which says that an arbitrarily long and accurate quantum computation is possible if the amount of decoherence per elementary quantum gate is below a certain value—the error threshold. Furthermore, the operational and spatial costs of fault-tolerance are benign in the sense of scaling, albeit large in absolute terms.

After the threshold theorem was established, work on fault-tolerant quantum computation has focused on (i) relaxing the preconditions under which the theorem can be applied \cite{FTQCLongRangeCorrelatedNoise, FTQCLocalNonMarkovianNoise}, (ii) improving the value of the error threshold \cite{Knill_2005}, (iii) reducing the cost of fault-tolerance \cite{Bravi_Kitaev_magic_distillation, gottesman2014faulttolerantquantumcomputationconstant, PantelevKalachev_LDPC, Wills_ConstantOverheadMagic, guernut2024faulttolerantconstantdepthcliffordgates}, and (iv) adapting fault-tolerance to specific architectures with certain operational constraints, such as limitations in the range \cite{raussendorf_topological_2007} or speed \cite{MonroeKim_IonTrap} of interaction.

The present work contributes to reducing the cost of fault tolerance, for a scheme of fault-tolerant quantum computation with high threshold that only requires nearest-neighbor entangling gates in a 3D or quasi-2D geometry. In this paper, we present a rigorous and complete description of Measurement-Based Quantum Computation in the RHG 3D cluster state \cite{raussendorf_topological_2007}. We first state and prove a general theorem on the topological implementation of gates by braiding holes in a Toric code. The corresponding tangle supports correlation surfaces that carry quantum information. The theorem is a three-dimensional analog of the foundational result of~\cite{raussendorf_measurement-based_2003}, which exhibits the intrinsic relation between the realization of a logical gate and the measurement pattern inside the 3D cluster. The theorem is flexible enough to allow for physical defects in the a priori regular and self-dual lattice supporting the cluster state. Using this, we show that the introduction of a local dislocation defect, which breaks the duality, allows for an intrinsically fault-tolerant implementation of the Hadamard gate, which was so far a missing gate in this setup. The dislocation defect may be regarded as a cluster state counterpart to the twist-defect in the Toric code \cite{Bombin_2010, Kitaev_2012}. However, in the present case, topological gates are obtained by braiding hole-type defects with the new dislocation defect, not by braiding twist defects among themselves. This and the Rudolph-Grover rebit encoding allow, in turn, for a topological implementation of the $S$ gate, yielding an intrinsically fault-tolerant, distillation-free, realization of the full Clifford group.

Altogether, these novelties in the setup of MBQC yield a gain of one order of magnitude in the overhead of the $S$ gate, compared to previous proposals involving magic state distillation. The final gate needed to achieve universality, the $T$ gate, remains in need of distillation. We propose an approach towards reducing the overhead using both topological and  geometric optimization of the circuit, with the goal of simplifying and compacting the tangle needed to implement the gates. Since implemented gates are given by surfaces supported on the tangle, the validity of the tangle is hard to verify by hand: we present a computer program that does this automatically throughout the circuit. For the concrete example of the Reed-Muller encoding used in the magic state distillation procedure required for a $T$ gate, our proposal has been verified by our program and reduces the overhead by half an order of magnitude.

The paper is organized as follows. Sec.~\ref{Sec:Cluster} first reviews the structure of the 3D cluster state as introduced in \cite{raussendorf_fault-tolerant_2006, raussendorf_topological_2007} and the basics of MBQC: how individually measuring the qubits in the bulk projects a Toric code on its boundary. Sec.~\ref{sec:MBQC THeorem} introduces the notions of measurement patterns and surfaces compatible with them. This allows for the proof of Theorem~\ref{Thm: Fundamental} which connects measurement patterns with unitary gates. After briefly reviewing elementary gates, Sec.~\ref{sec:clifford + T gate} discusses the Hadamard gate, the $S$ gate, and analyzes their respective overhead. In Sec.~\ref{Sec: Performance}, we concentrate on the $T$ gate and compute the associated overhead. Finally, we turn to circuit volume optimization in Sec. \ref{subsec: circuit volume optimization}. We propose equivalent circuits for magic state distillation, focusing on the improvement over a `naive' implementation. We additionally developed and implemented an algorithm able to verify the optimization: It is described in the appendix and available at \cite{Schwiering2025}.

\section{The 3D cluster state and the Toric code}\label{Sec:Cluster}

\begin{figure}[h!]
	\centering
    \begin{subfigure}{0.35\textwidth}
	\includegraphics[width=1\textwidth]{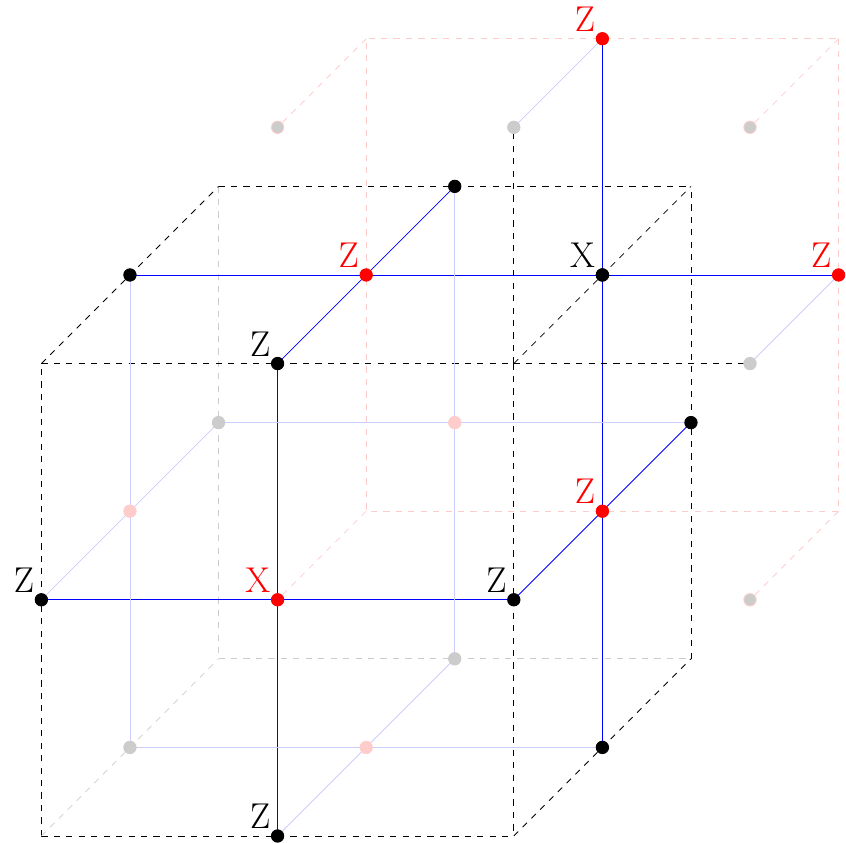}
    \caption{Unit cell of the 3D cluster state. }
    \end{subfigure}
    \qquad
    \begin{subfigure}{0.55\textwidth}
	\includegraphics[width=1\textwidth]{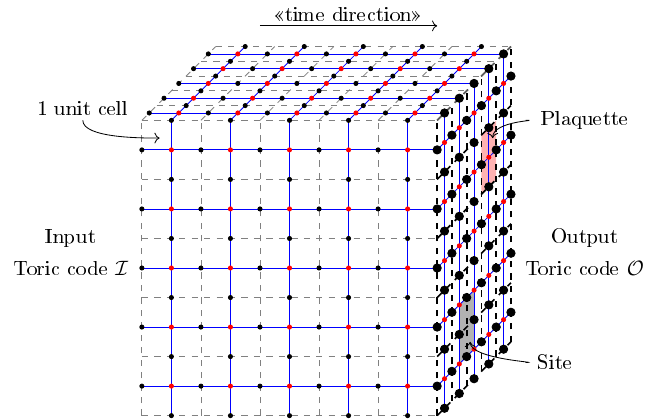}
    \caption{$5\times 5\times 5$ cluster state lattice.}
    \end{subfigure}
    \caption{The cluster state lattice: one unit cell (a) and a $5\times 5\times 5$ lattice (b). The physical qubits are located on the red and black dots. The entanglement induced by the $CZ$ gates is represented as solid blue lines. The underlying cubic lattice is represented by dashed lines, black qubits sit on the edges, red qubits on the faces. (b) For computation purposes, the input Toric code $\mathcal{I}$ is inserted on the cluster left plane (not visible on the figure because of perspective). It is foliated in the horizontal direction, thus simulating a \guillemotleft time\guillemotright ~evolution. The output Toric code $\mathcal{O}$ is made of the highlighted black qubits on the right plane. }
	\label{fig: Cluste state unit cell}
\end{figure}

We start with a brief review of the three-dimensional cluster state \cite{briegel2001persistent} and its relation to the two-dimensional Toric code \cite{dennis_topological_2002, kitaev2003fault}. Following~\cite{raussendorf_topological_2007}, physical qubits are placed on the edges of two simple cubic lattices (which we refer to as primal, respectively dual) that are shifted from each other so that the centers of the primal cubes are the corners of the dual ones and vice versa, see Fig.~\ref{fig: Cluste state unit cell}, where we distinguish the two lattices by a black, respectively red color. The fundamental graph defining the cluster state (in blue in the figure)  is made up of nearest-neighbor qubits. To prepare the cluster state, one performs a $CZ$ gate on every edge of the graph with all qubits initially in  the $\ket{+}$ state. Note that these gates all commute with each other and that $CZ_{a,b}\ket{++}=CZ_{b,a}\ket{++}$. Therefore, the unambiguously defined resulting global state is given by 
\begin{equation}
	\left| \phi \right\rangle_{\mathcal{C}} = \prod_{\left\langle r,b \right\rangle }	CZ_{r,b} \bigotimes_{q\in \mathcal{C} } \left| + \right\rangle_q
	\label{eq: construction of the cluster state}
\end{equation}
where $\langle r,b \rangle$ stands for nearest neighbors (which are necessarily of different color) and $\mathcal{C}$ denotes the set of sites carrying a qubit.

We recall the homological language introduced in~\cite{raussendorf_topological_2007}, which best describes the mathematical structure underlying the cluster state, see also~\cite{kitaev2003fault,bachmann_local_2017}. This language has recently received increasing interest, in particular in the context of Low Density Parity Check codes \cite{rakovszky2024physicsgoodldpccodes}. The primal cubes and dual cubes are in fact the three-dimensional cells of two cell complexes that are dual to each other. We denote the set of primal volumes by~$G_3$, of primal square by~$G_2$, of primal edges by~$G_1$ and of primal vertices by~$G_0$. The $n$-chain vector spaces will be denoted by $C_n = C_n(G_n,\F_2)$ and it is made of formal linear combinations of $n$-cells with $\F_2$ coefficients, namely $c = \sum_{g\in G_n} \alpha_g g$ with $\alpha_g\in\F_2 = \{0, 1\}$. We denote the usual boundary operator $\partial_n: C_n\to C_{n-1}$. The same can be done with the dual complex, which we denote $\overline{C}_3, \overline{C}_2, \overline{C}_1, \overline{C}_0$. The physical qubits lie on primal and dual edges. The two complexes are dual to each other with the duality given by
\begin{equation} \label{dual chain space}
    \overline C_n \,\cong\, C_{3-n}
\end{equation}

Since
\begin{equation}\label{X CZ commutation}
    X_a Z_b CZ_{a,b} = CZ_{a,b}X_a,
\end{equation}
a cluster state can equivalently be characterized by the fact that its stabilizer group is generated by the local operators
\begin{equation*}
    K(a)= X_a \bigotimes_{b\in \partial a} Z_b
\end{equation*}
where $a\in G_2$ or $a\in\overline G_2$, see again Fig.~\ref{fig: Cluste state unit cell}. We further note that $K(a)^2 = \mathbb{I}$ and that this is a set of commuting operators. Consequently, the stabilizer group $\mathcal K$ of the cluster state is isomorphic to the group $C_2\oplus\overline C_2$.

This isomorphism can be made explicit by the following expression for a general stabilizer in terms of Pauli operators. For any $c_2 \in C_2$, we define the following stabilizer:
\begin{equation*}
    K(c_2)= \prod_{a\in G_2}K(a)^{\alpha_a}.
\end{equation*}
Since all edges that are not on the boundary $\partial c_2$ appear twice in the product and $Z^2 = \mathbb{I}$, the stabilizer has the simpler expression
\begin{equation}
    K(c_2)= X(c_2) Z(\partial c_2),
    \label{eq: stabilizers on  surfaces}
\end{equation}
where $X(c_2)= \bigotimes_{r\in c_2} X_r$ and similarly for $Z(\partial c_2)$. For later consistency, we also set $X(\emptyset) = \mathbb{I} = Z(\emptyset)$. In other words, this is the product of $X$ on the red qubits inside the surface times a product of $Z$ on the black qubits on the boundary. The same holds of course for dual surfaces and yield 
\begin{equation}
	K(\overline{c}_2)= X(\overline{c}_2) Z(\partial \overline{c}_2).
	\label{eq: stabilizers on dual surfaces}
\end{equation}

Some care must be taken at the boundary of the cluster, where the surfaces may end. For simplicity, we consider again Fig.~\ref{fig: Cluste state unit cell} and imagine that the primal face lies on the boundary of the cluster. Then the dual cube sketched is only partially in the cluster, and one checks that the product $X$'s of all five primal qubits is a stabilizer. This is consistent with~(\ref{eq: stabilizers on dual surfaces}) under the convention that the truncated faces are bonafide dual faces. The same holds, of course, with dual objects.

With this in hand, we are ready to connect the 3D cluster state to the 2D Toric code. We consider the cluster state on a large cube bounded by primal surfaces (see for instance Fig.~\ref{fig: Cluste state unit cell}b for a $5\times 5\times 5$ lattice) and perform $X$-measurements on all qubits but those on the very last plane (output Toric code $\mathcal{O}$ in Fig.~\ref{fig: Cluste state unit cell}), on which we only measure the dual qubits (red qubits) in $X$ as well. The resulting state is a product of an entangled state on the primal qubits on $\mathcal{O}$ with a random product state of $\ket{\pm}$ on the rest of the cluster --- the bulk of the cluster has been `reset'. By the above discussion, it is stabilized exactly by the Toric code stabilizers up to a sign given by the measurement outcomes. In other words, the state resulting from the measurements in the bulk is an eigenstate of the Toric code Hamiltonian, the eigenvalue being determined by the result of all measurements. This correspondence is central to the realization of topological Measurement-Based Quantum Computation. Note that here, the qubits on the input Toric code $\mathcal{I}$ in Fig.~\ref{fig: Cluste state unit cell}b are also measured in the $X$ basis. We don't perform any computation here but just prepare a Toric code from a cluster state. The corresponding plaquette and site have been highlighted in red and black respectively. 

Before delving further into this, we note that although this measurement-based procedure for creating a topologically ordered state has been known since the original~\cite{raussendorf_topological_2007}, the idea was recently revived~\cite{GaugingMeasurement} and used to prepare even non-Abelian topological orders experimentally~\cite{tantivasadakarnExp}.

\section{Measurement-Based Quantum Computation in the cluster state}
\label{sec:MBQC THeorem}
The cluster state being obtained by a finite depth quantum circuit, it is short-range entangled, but in a non-trivial Symmetry Protected Topological phase. Local measurements then realize the gauging to a proper topological order (see also~\cite{GaugingMeasurement} for a general point of view on this), which is the fundamental process enabling MBQC. 

\subsection{Long-range entanglement through measurements}

\begin{figure}[h!]
	\centering
	\includegraphics[width=0.5\textwidth]{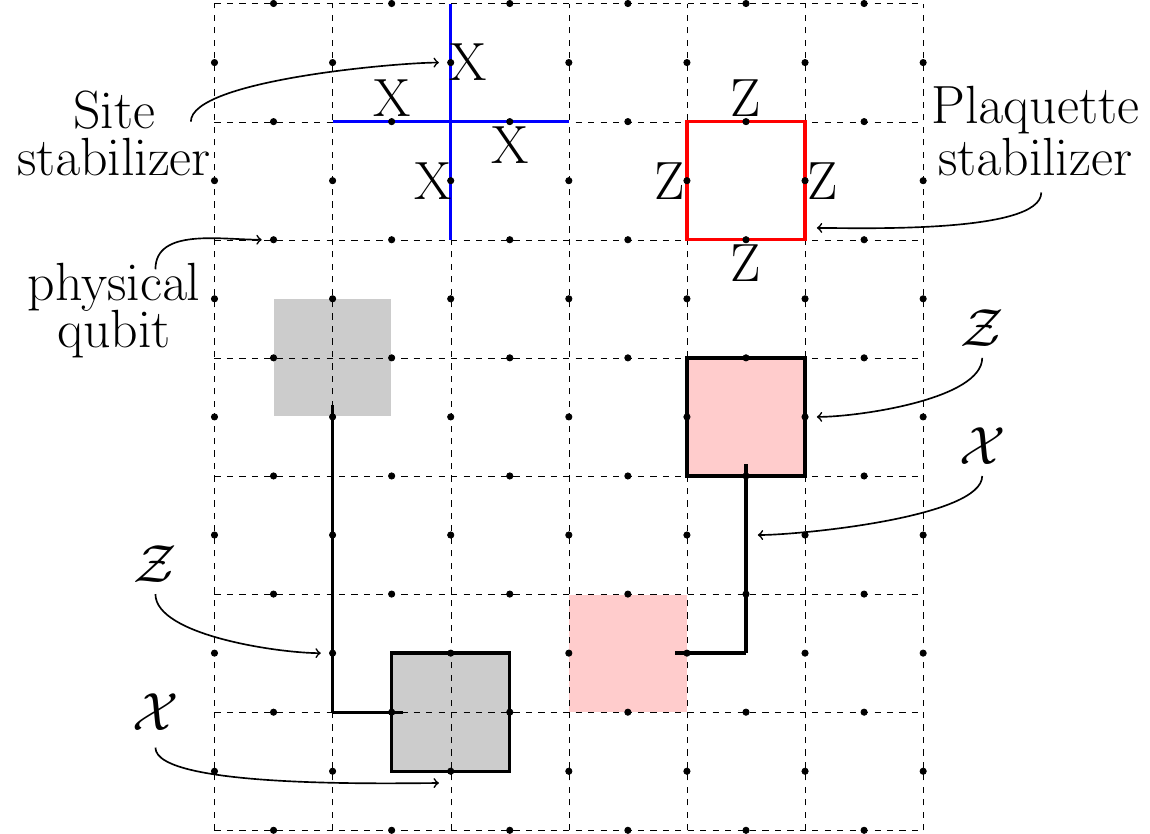}
    \caption{The Toric code. The physical qubits sit on the graph edges, represented by small black dots.  At the top are defined two stabilizers of different types: plaquette Z and site X. Logical qubits are encoded in pairs of missing stabilizers (blue and red squares at the bottom), they can be of two types due to the duality of the surface code. A logical Pauli operator correspond to a product of Pauli operators on physical qubits along a string either extending from one hole to the other or winding around one hole.}
	\label{fig:Toric code figure}
\end{figure}

Information processing in MBQC is performed by teleportation from plane to plane in the cluster. We recall how qubits can be encoded in a Toric code. The fault-tolerant encoding is obtained by punching holes in the code: indeed, to any pair of stabilizers of the same type that is removed from the set of generators of the stabilizer group (a pair of `holes') corresponds a pair of operators $\logicX,\logicZ$ acting non-trivially on the code and satisfying the commutation relations of the Pauli matrices. In the case of two removed plaquette stabilizers, $\logicX$ is given by the product of $X$ operators along a string extending from one hole to the other while $\logicZ$ is a loop winding once around the plaquette, see Fig.~\ref{fig:Toric code figure}. Since closed string operators act trivially on the code, all strings extending from one hole to the other and all strings winding around the hole are equal when restricted to the code space. Since $\logicX,\logicZ$ are in the center of the stabilizer group, they are bonafide operators acting on the code space.

\begin{figure}[h!]
	\centering
	\includegraphics[width=0.45\textwidth]{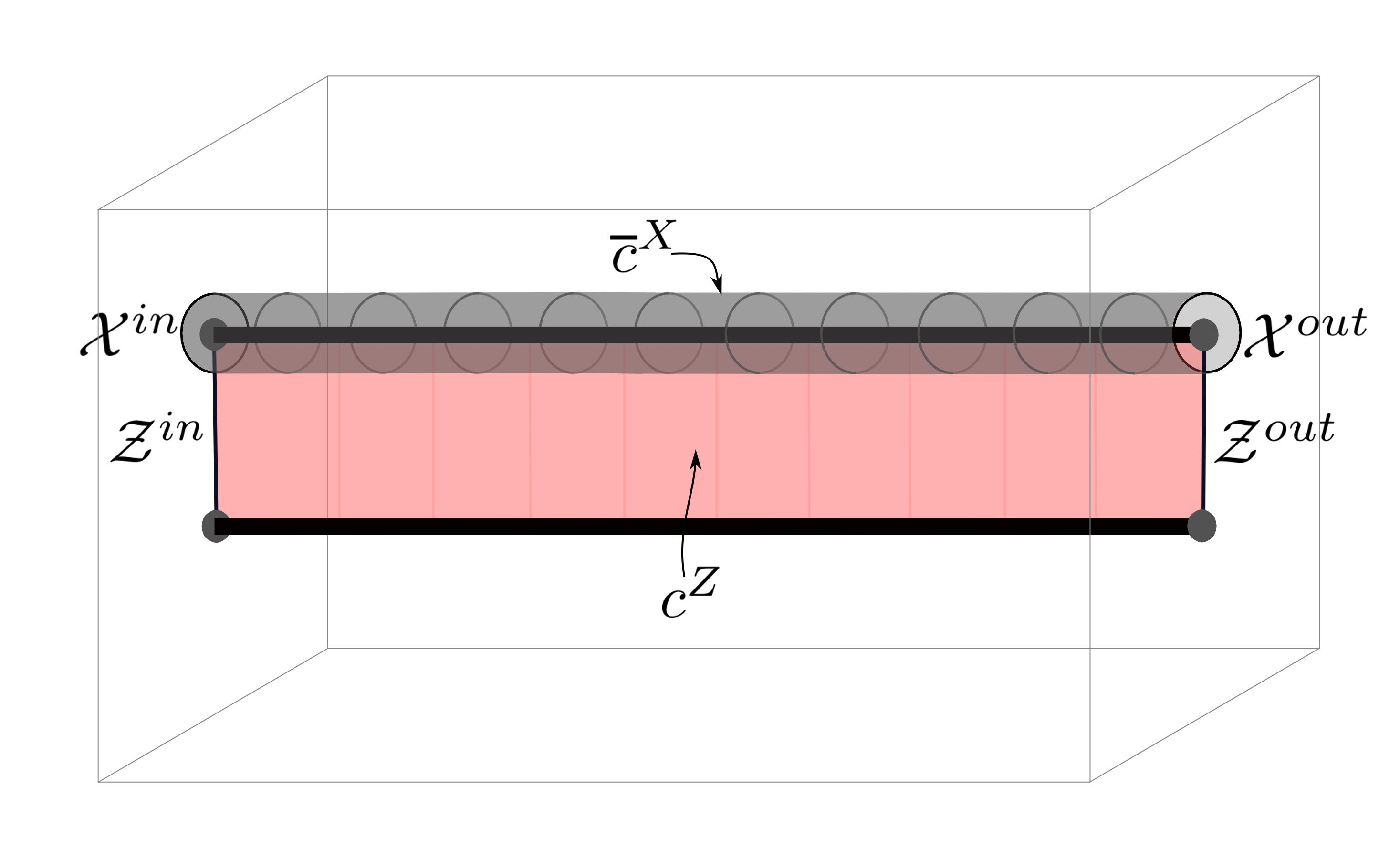}
    \caption{The surfaces that correspond to the identity gate. The surface $c^Z$ is primal with black qubits on its boundary, while $\bar c^X$ is dual and has no boundaries at all. The black qubits in $\partial c^Z\cap \mathcal{B}$  are measured in $Z$, while all the other qubits in $\mathcal{B}$ are measured in $X$. After the measurements in the bulk, the state on $\mathcal{I}\cup \mathcal{O}$ is an encoded Bell pair. }
    \label{fig:Identity}
\end{figure}

This picture motivates the following. We consider a large cluster with two distinguished opposite planes, which we call the `in' and `out' planes. The qubits in the cluster are divided into three sets
\begin{equation*}
    \mathcal{C} = \mathcal{I}\cup\mathcal{B}\cup\mathcal{O}
\end{equation*}
corresponding to the `in' plane $\mathcal{I}$, the `out' plane $\mathcal{O}$ and the bulk $\mathcal{B}$. Let $c^Z$ be a primal surface that is a rectangle having one side on each of $\mathcal{I},\mathcal{O}$ while the other two extend in its bulk, see Fig.~\ref{fig:Identity}. We modify the measurement procedure described in Section~\ref{Sec:Cluster}, firstly by not measuring the qubits in $\mathcal{I},\mathcal{O}$, and secondly by measuring the qubits on the inner boundary of the surface, $\partial c_2\cap \mathcal{B}$, in the $Z$ basis while all the other qubits in the bulk, including the ones inside the surface $c^Z$, are measured in $X$. Since $K(c^Z)$ is a stabilizer that commutes with the measurements, we conclude that the final state is an eigenstate of $\logicZ_{\mathrm{in}}\otimes \logicZ_{\mathrm{out}}$ where $\logicZ_{\mathrm{in}} = Z(\partial c^Z\cap \mathcal{I})$ and similarly for $\logicZ_{\mathrm{out}}$. Furthermore, let $\bar c^X$ be a boundaryless cylinder surrounding one of the edges of the rectangle extending from the initial to the final plane, see Fig.~\ref{fig:Identity}. The corresponding stabilizer $K(\bar c^X)$ is a product of $X$ operators on the cylinder that commute with the $X$ measurements in the bulk. Note that this surface cannot be contracted as it wraps around a line of qubits measured in $Z$. The state after the measurements in the bulk is also an eigenstate of $\logicX_{\mathrm{in}}\otimes \logicX_{\mathrm{out}}$. Together, we have therefore obtained an \emph{encoded Bell pair} across the cluster. The fundamental theorem of topological MBQC, see~Theorem~\ref{Thm: Fundamental} below, states that an appropriate choice of measurements, which correspond to primal and dual surfaces, implements a unitary between the encoded initial state and the encoded final state.

\subsection{A dynamical point of view}

This `global' picture can also be described in local, `time dependent' way as follows. Initially, logical qubits are encoded in an eigenstate $\ket{\varphi_{\mathrm{in}}}$ of a Toric code. Then the physical qubits in the Toric code are entangled to a cluster state via $CZ$ gates so that the qubits in the Toric code become the black qubits on $\mathcal{I}$. The state obtained by this coupling is not exactly a cluster state because the physical qubits in the Toric code were not individually in the $\ket+$ state.

A measurement in $\mathcal{B}$ corresponds to the projector $P_\mathcal{B}(\underline s,\underline\sigma) = \bigotimes_{q \in \mathcal{B} }  \left( \frac{1 + s_q \sigma_q}{\sqrt2} \right)$, where $\underline\sigma = \{\sigma_q:q\in\mathcal{B}\}$ is a vector of (possibly rotated) Pauli operators and $\underline s = \{s_q:q\in\mathcal{B}\}$ is the $\{0,1\}$-valued vector containing the measurement outcomes. The resulting state on the entire cluster $\mathcal{C}$ is given by
\begin{equation}
	\left| \psi \right\rangle = P_\mathcal{B}(\underline s,\underline\sigma) \prod_{\left\langle r,b \right\rangle }	CZ_{r,b} \Big(\ket{\varphi_{\mathrm{in}}} \bigotimes_{q\in \mathcal{B} \cup \mathcal{O} } \left| + \right\rangle_q \Big) 
	\label{eq: state after measurement}
\end{equation}
In practice, all qubits will be measured in either the $X$ or the $Z$ basis, with a few isolated exceptions measured in the $X\!-\!Y$ plane that we shall use for magic state distillation. Here again, a final measurement of the qubits on the initial plane yields a product state between $\mathcal{I}$, $\mathcal{B}$ and $\mathcal{O}$; We denote the latter part, which is a Toric code state, by $\ket{\varphi_{\mathrm{out}}}$.

Given a $P_\mathcal{B}(\underline s,\underline\sigma)$, an element $c_2\in C_2$ (or $\overline c_2\in C_2$) is said to be \emph{compatible} with the measurement if
\begin{equation}\label{Compatible surface}
    [K(c_2),P_\mathcal{B}(\underline s,\underline\sigma)] = 0.
\end{equation}
Concretely, if
\begin{equation*}
   \begin{cases}
       &\sigma_q = X_q \quad\text{whenever}\quad q\in c_2 \\
       &\sigma_q = Z_q \quad\text{whenever}\quad q\in \partial c_2
   \end{cases}
\end{equation*}
then $c_2$ is compatible with $P_\mathcal{B}(\underline s,\underline\sigma)$ (the same holds for dual $2$-chains).

The following theorem, which generalizes~\cite{raussendorf_fault-tolerant_2006,raussendorf_topological_2007}, establishes the connection between the measurement pattern in the bulk of the cluster and the information processed between the input Toric code to an output layer. Its proof follows a reasoning similar to the two-dimensional argument of Theorem~1 in~\cite{raussendorf_measurement-based_2003}, but adapted to the 3D case. The idea is simple: given a measurement pattern, one searches for surfaces compatible with it. They in turn correspond to logical qubits on the input and output layers. By simply following the surfaces from the input qubits to the output qubits, one reads off transformation laws mapping elements of the real Pauli group to themselves, namely: to a set of compatible surfaces corresponds a real Clifford gate. We recall that Clifford gates are all unitary elements of the normalizer of the Pauli group. The \emph{real Clifford gates} are the unitary elements of the normalizer of the real Pauli group, namely of the group generated by $X$ and $Z$ only.

The setting is as follows. The initial and final Toric codes come with missing stabilizers and corresponding logical operators $\{\logicX_i^{\mathrm{in}},\logicZ_i^{\mathrm{in}}: i=1,\ldots n\}$ and $\{\logicX_i^{\mathrm{out}},\logicZ_i^{\mathrm{out}}: i=1,\ldots n\}$. The initial state $\ket{\varphi_{\mathrm{in}}}$ is now entangled with the rest of the cluster. Measurements in the bulk are given and described by the projection $P_\mathcal{B}(\underline s,\underline\sigma)$, yielding the state~(\ref{eq: state after measurement}). A final measurement is carried out on the initial plane, in the $X$ basis, which we describe by the projector $P_\mathcal{I}(\underline t,\underline X)$. Slightly abusing notation, we denote $\ket{\varphi_{\mathrm{out}}}$ both the state of the full cluster and the state on the final Toric code. The assumptions of the theorem ensure not only that surfaces are compatible with the measurements, but also with the encoded qubits of the initial and final codes.

\begin{theorem}[Fundamental Theorem of Topological MBQC] \label{Thm: Fundamental} 
Let $U$ be a real Clifford gate. Assume that for each $i\in\{1,\ldots,n\}$, there are $c^X_i, c^Z_i\in C_2$ and $\bar c^X_i,\bar c^Z_i\in \overline C_2$ that are all compatible with the measurements $P_\mathcal{B}(\underline s,\underline\sigma)$ and such that $\partial c^X_i \cap \mathcal{I} = \emptyset$ and $\bar c^Z_i \cap \mathcal{I} = \emptyset$ and
\begin{align}
    & X(\overline c^X_i\cap \mathcal{I}) = \logicX_i^{\mathrm{in}} \label{eq: X condition on IN surface}\\
    & X(\overline c^X_i\cap \mathcal{O})Z(\partial c^X_i \cap \mathcal{O}) = U \logicX_i^{\mathrm{out}} U^\dagger
    \label{eq: X condition on OUT surface}
\end{align}
and 
\begin{align}
    & Z(\partial c^Z_i\cap \mathcal{I}) = \logicZ_i^{\mathrm{in}} \label{eq: Z condition on IN surface}\\
    & X(\overline c^Z_i\cap \mathcal{O})Z(\partial c^Z_i \cap \mathcal{O}) = U \logicZ_i^{\mathrm{out}} U^\dagger.
    \label{eq: Z condition on OUT surface}
\end{align}
Let $\ket{\varphi_{\mathrm{out}}} = P_\mathcal{I}(\underline t,\underline X)\ket{\psi}$. Then
\begin{equation*}
	\ket{\varphi_{\mathrm{out}}}  = U U_{P} \ket{\varphi_{\mathrm{in}}},
\end{equation*}
where $	U_{P} = \mathrm{e}^{\mathrm{i}\eta}\bigotimes_{i = 1 }^{n} (\logicX_i^{\mathrm{out}})^{\underline s_i^Z}  (\logicZ_i^{\mathrm{out}})^{\underline t_i^X+\underline s_i^X}$. Here, $s_i^Z$ are the measurement outcomes in $c_i^Z$ and similarly for the other exponents.
\end{theorem}

We point out that we assumed the initial and final planes $\mathcal{I}$ and $\mathcal{O}$ that define the input and output Toric codes are made of primal edges (black qubits on Fig.~\ref{fig: Cluste state unit cell}) only, which implies that
\begin{equation}\label{Primal In and Out planes}
    \partial \bar c_i^{X,Z}\cap (\mathcal{I}\cup\mathcal{O}) = \emptyset,\qquad c_i^{X,Z}\cap (\mathcal{I}\cup\mathcal{O}) = \emptyset.
\end{equation}
From the cluster point of view, the red qubits belonging to these planes are part of the measured bulk $\mathcal{B}$ and are therefore measured according to the measurement pattern.

\begin{proof}
We first prove the theorem in the case where all logical qubits of the input are in the logical $+$ state, namely
\begin{equation}\label{logical + in state}
\logicX_i^{\mathrm{in}}\ket{\varphi_{\mathrm{in}}^+} = \ket{\varphi_{\mathrm{in}}^+}
\end{equation}
for all $i=1,\ldots,n$. We denote $\ket{\psi (+)}$ the corresponding state~(\ref{eq: state after measurement}) after the bulk measurement. Since~(\ref{eq: X condition on IN surface}) implies that $X(\overline{c}_i^X\cap \mathcal{I})\ket{\varphi_{\mathrm{in}}^+} = \logicX_i^{\mathrm{in}}\ket{\varphi_{\mathrm{in}}^+}=\ket{\varphi_{\mathrm{in}}^+}$ and the other physical qubits are eigenstate of the local $X$ Pauli operators, we have that
\begin{equation*}
	X(c_i^X)X(\overline{c}_i^X) \Big(\ket{\varphi_{\mathrm{in}}^+} \bigotimes_{q\in \mathcal{B} \cup \mathcal{O} } \left| + \right\rangle_q \Big) =\Big(\ket{\varphi_{\mathrm{in}}^+} \bigotimes_{q\in \mathcal{B} \cup \mathcal{O} } \left| + \right\rangle_q \Big)  
\end{equation*}

We now use~(\ref{X CZ commutation}) to commute the $X$ operators through the $CZ$ in~$\ket{\psi (+)}$, resulting in
\begin{equation}\label{Compatible single qubit X surface}
    \prod_{\left\langle r,b \right\rangle }	CZ_{r,b} X(c_i^X)X(\overline{c}_i^X)
    =K(c_i^X) K(\overline{c}_i^X) \prod_{\left\langle r,b \right\rangle }	CZ_{r,b}
\end{equation}
Since the surfaces are compatible with the measurements~(\ref{Compatible surface}), the operators commute with $P_\mathcal{B} (\underline s,\underline\sigma)$ and so
\begin{align}
    \ket{\psi (+)}&=K(c_i^X) K(\overline{c}_i^X)\ket{\psi (+)} \nonumber \\
	&= (-1)^{\underline s_i^X} X(\overline{c}_i^X\cap \mathcal{I}) X(\overline{c}_i^X\cap \mathcal{O})Z(\partial c_i^X\cap \mathcal{O}) \left| \psi (+)\right\rangle \nonumber\\
	&=(-1)^{\underline s_i^X} \logicX_i^{\mathrm{in}} \otimes (U \logicX_i^{\mathrm{out}} U^\dagger ) \left| \psi (+)\right\rangle
 \label{eq: psi + before in measurement}
\end{align}
where we used $\partial c^X_i \cap \mathcal{I} = \emptyset$ and~(\ref{eq: X condition on IN surface},\ref{eq: X condition on OUT surface}) in the last equality. Here, $(-1)^{\underline s_i^X}$ is the result of all measurements on the $i$th surface. After the measurements in the initial plane, the state is given by
\begin{equation}
P_\mathcal{I}(\underline t,\underline X)\ket{\psi (+)}
=(-1)^{\underline t^X+\underline s^X}\ket{\underline t_i}\otimes \ket{\underline s_i} \otimes U \logicX_i^{\mathrm{out}} U^\dagger \ket{\varphi_{\mathrm{out}}^+}
	\label{eq: eigen eq state out +}
\end{equation}
Since this holds for all $i=1,\ldots,n$, we conclude that $ \ket{\varphi_{\mathrm{out}}^+}$ is an eigenstate of $U \logicX_i^{\mathrm{out}} U^\dagger$ for all $i$, namely
\begin{equation}
    U^\dagger \ket{\varphi_{\mathrm{out}}^+} = (-1)^{\underline t_i^X+\underline s_i^X} \logicX_i^{\mathrm{out}} U^\dagger \ket{\varphi_{\mathrm{out}}^+}.
    \label{X eigenvalue equation}
\end{equation}
In other words, $U^\dagger \ket{\varphi_{\mathrm{out}}^+}$ is, up to a phase $\mathrm{e}^{\mathrm{i}\eta(+)}$, in the logical $\pm$ state for each $i$. The theorem follows in this special case by comparing this with~(\ref{logical + in state}) and observing that the necessity or not of a spin flip $\logicZ_i^{\mathrm{out}}$ is determined by the corresponding eigenvalue.

We now consider the case of an input state where the qubits are all in the logical $Z$ basis, namely
\begin{equation}\label{logical Z in state}
\logicZ_i^{\mathrm{in}}\ket{\varphi_{\mathrm{in}}^{\underline z}} = (-1)^{z_i}\ket{\varphi_{\mathrm{in}}^{\underline z}}
\end{equation}
If $P(\underline z) = \ket{\varphi_{\mathrm{in}}^{\underline z}}\bra{\varphi_{\mathrm{in}}^{\underline z}}$, then $\ket{\varphi_{\mathrm{in}}^{\underline z}} = 2^{n/2}P(\underline z)\ket{\varphi_{\mathrm{in}}^+}$. Moreover, $P(\underline z)$ commutes with any $CZ$ gate, and so the state~(\ref{eq: state after measurement}) after the measurements in the bulk is given by
\begin{equation}
    \ket{\psi(\underline z)} = 2^{n/2} P(\underline z)\ket{\psi(+)}
\end{equation}

We repeat the argument above, but with surfaces $c_i^Z,\bar c_i^Z$, which yields~(\ref{Compatible single qubit X surface}) for the corresponding stabilizer $K(c_i^Z)K(\bar c_i^Z)$. Since $\bar c^Z_i \cap \mathcal{I} = \emptyset$, the only operator acting on the initial surface is $Z(\partial c_i^Z\cap\mathcal{I})$, which commutes with $P(\underline z)$. Hence,
\begin{align*}
    \ket{\psi(\underline z)} 
    &= 2^{n/2} (-1)^{\underline s_i^Z}  P(\underline z) \logicZ_i^{\mathrm{in}} X(\overline{c}_i^Z\cap \mathcal{O})Z(\partial c_i^Z\cap \mathcal{O}) \ket{\psi(+)} \\
    &=(-1)^{\underline s_i^Z+z_i} X(\overline{c}_i^Z\cap \mathcal{O})Z(\partial c_i^Z\cap \mathcal{O}) \ket{\psi(\underline z)}.
\end{align*}
The state on the final plane is not entangled with the the qubits on the initial plane so that the measurement on $\mathcal{I}$ does not affect it. Using~(\ref{eq: Z condition on OUT surface}), we obtain the equation
\begin{equation}\label{Final z}
    U^\dagger \ket{\varphi_{\mathrm{out}}^{\underline z}} = (-1)^{\underline s_i^Z+z_i} \logicZ_i^{\mathrm{out}} U^\dagger \ket{\varphi_{\mathrm{out}}^{\underline z}}.
\end{equation}
Hence $U^\dagger \ket{\varphi_{\mathrm{out}}^{\underline z}}$ is again a computational basis state with some logical qubits that are flipped as compared with the initial state~(\ref{logical Z in state}). This is again the statement of the theorem in these cases, with a phase $\mathrm{e}^{\mathrm{i}\eta(\underline z)}$. Note that here $z_i$ is not due to measurements but is given by the initial state of the logical qubit $i$; Since it appears both in~(\ref{Final z}) and~(\ref{logical Z in state}), the necessity of $\logicX_i^{\mathrm{out}}$ is determined by $\underline s_i^Z$.

It remains to decompose $\ket{\varphi_{\mathrm{out}}^{+}}$ in the basis $\ket{\varphi_{\mathrm{out}}^{\underline z}}$ to conclude that all phases $\eta(\underline z)$ are equal to $\eta(+)$, see~\cite{raussendorf_measurement-based_2003}. This determines $U_P$ (up to a global phase) on the basis $\ket{\varphi_{\mathrm{in}}^{\underline z}}$, and thus by linearity on any input vector $\ket{\varphi_{\mathrm{in}}}$, concluding the proof.
\end{proof}

Note that $U$ is completely determined by the measurement in $\mathcal{B}$ and the existence of primal and dual surfaces that `connect' the Pauli operators in $\mathcal{I}$ and $\mathcal{O}$: Indeed, if such surfaces exist, after the measurements in $\mathcal{I}$, the state $\ket{\varphi_{\mathrm{out}}}$ is defined uniquely and so is $U$. Therefore, in order to implement a given Clifford gate $U$, one must find a suitable measurement pattern: Since this pattern is not unique, there is the possibility of optimization to reduce the size of the cluster $\mathcal{B}$. This will be the main focus of Section~\ref{subsec: circuit volume optimization} for the case of multiple $\CNOT$ gates. 

\subsection{Illustrative examples: CSS gates}
\label{subsec: CSS gates}
\begin{figure}[h!]
	\centering
	\includegraphics[width=0.45\textwidth]{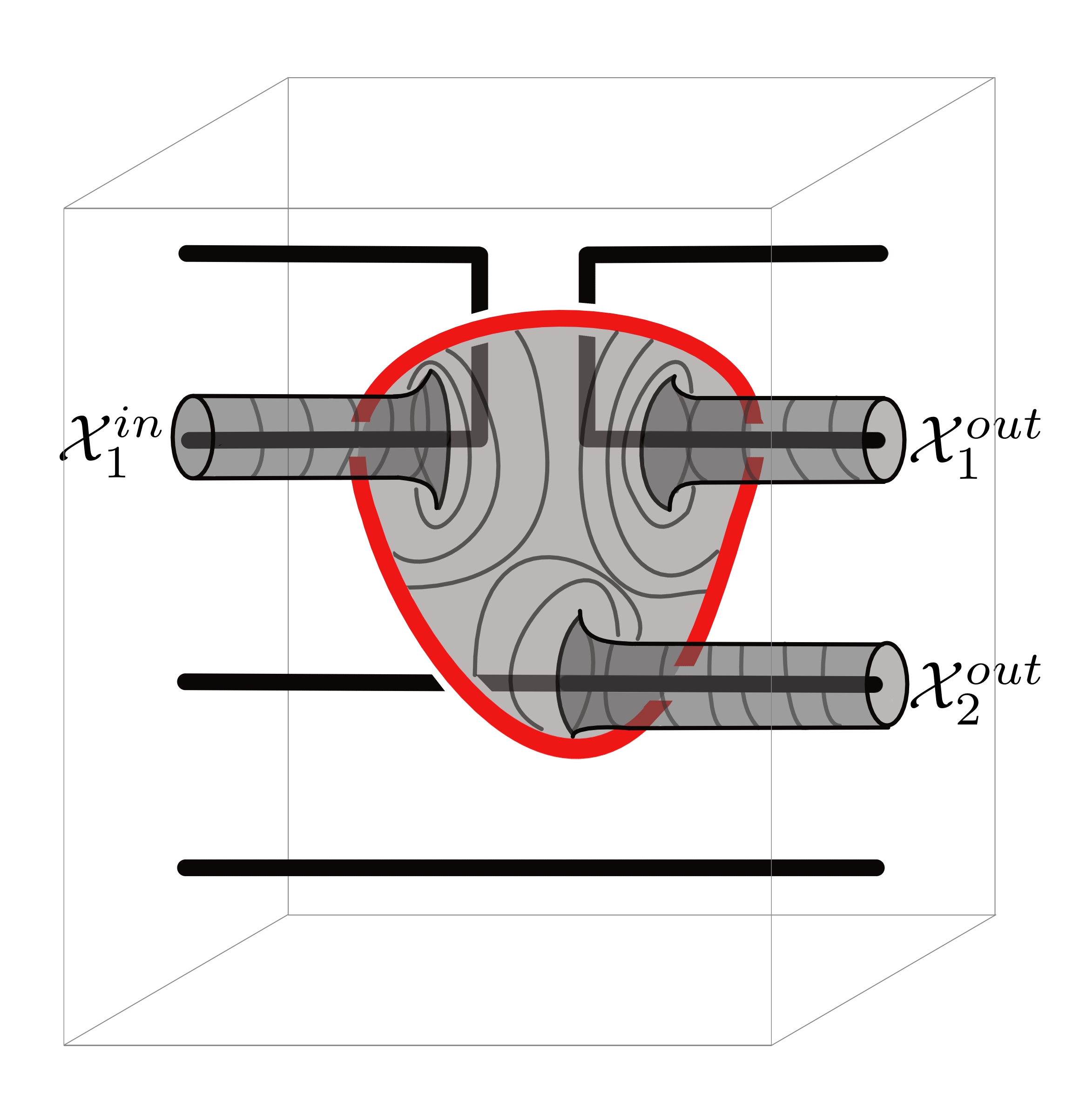}
    \caption{The input and output qubits and the measurement lines that correspond to the $\CNOT$ gate. The gray surface sketch the `tubes' making up the $\bar c_1^X$ surface, while the red line is its boundary, which is made up of dual qubits. }
	\label{fig: CNot}
\end{figure}

Here we briefly review the measurement patterns proposed in~\cite{raussendorf_topological_2007} to implement the identity and the $\CNOT$ (see Fig.~\ref{fig:Identity} and \ref{fig: CNot}). With these, we can implement what we call \emph{CSS gates}, which map a logical $\logicX_i^{\mathrm{in}}$ to a product of logical $\logicX_i^{\mathrm{out}}$ and the same for $\logicZ$, in reference to the CSS codes \cite{CalderankShorPeter_Good_ECC}.

For the identity, a side view of the cluster state is presented in Fig.~\ref{fig:Identity} where the input Toric code is on the left and the output plane is on the right. The surfaces of Theorem~\ref{Thm: Fundamental} are: the primal $c^Z$ is plotted in gray and the dual $\bar c^X$ is represented in light red, while $c^X,\bar c^Z = \emptyset$. We immediately see that~(\ref{eq: X condition on IN surface})---(\ref{eq: Z condition on OUT surface}) are satisfied for $U = \mathbb{I}$.

The $\CNOT$ gate is slightly more complex as it entangles two logical qubits. In Fig.~\ref{fig: CNot}, we represent only the lines of $Z$ measurements: in black along primal boundaries and in red along dual boundaries. These lines form a non-trivial knot in three dimensions. The surfaces are here $c_1^Z,c_2^Z$ and $\bar c_1^X,\bar c_2^X$ and they connect the qubits on $\mathcal{I}$ to those on $\mathcal{O}$ as follows:
\begin{equation}
\begin{aligned}
    \bar c_1^X: & \: \logicX_1^{\mathrm{in}}\to\logicX_1^{\mathrm{out}}\logicX_2^{\mathrm{out}} && \text{(its boundary in $\mathcal{B}$ is the red line, see Fig.~\ref{fig: CNot})}\\
    \bar c_2^X: & \: \logicX_2^{\mathrm{in}}\to\logicX_2^{\mathrm{out}} \\
    c_1^Z: & \: \logicZ_1^{\mathrm{in}}\to\logicZ_1^{\mathrm{out}} && \text{(wraps around the `top' part of the red line)}\\
    c_2^Z: & \: \logicZ_1^{\mathrm{in}}\to\logicZ_1^{\mathrm{out}}\logicZ_2^{\mathrm{out}}
    && \text{(wraps around `lower right' part of the red line)}
\end{aligned}
\label{eq: surfaces CNOT}
\end{equation}
Comparing this with the statement of the theorem yields $U = \CNOT$.

\subsection{Comparison with topological MBQC on surface codes}

In this work, we focus on braiding hole defects to implement some computation on the Toric code. Another approach to topological MBQC (or topological circuit based QC in general) is to use one patch of surface code (same stabilizers as the Toric code but with open boundary conditions) to encode one logical qubit \cite{dennis_topological_2002}. In circuit-based computation, a  transversal logical $\CNOT$ gate can be implemented between two patches by applying a physical $\CNOT$ between a physical qubit in the first patch and its corresponding qubit in the second patch. A downside to this implementation is that every qubit in a patch needs to interact with another in the other patch, which could be practically hard to implement in an architecture where the patches were side by side. To resolve this issue, the notion of lattice surgery was introduced \cite{Horsman_2012_lattice_surgery}. Two patches are stitched together on one of their boundary; their logical operators merge. The patches are then separated and regain their individual logical operators, but they are now logically entangled. This procedure can be translated into an MBQC scheme quite naturally \cite{Herr_2018_lattice_surgery_mbqc}. 

In both frameworks, the Hadamard can be achieved topologically, and magic state distillation is used for universality. For instance, in circuit based, a transversal Hadamard is applied on every physical qubit of the patch followed by the patch rotation to recover the initial stabilizer structure\cite{Horsman_2012_lattice_surgery}. Patch rotation can be achieved by expanding the surface code lattice and applying local measurements. To translate this Hadamard into MBQC, \cite{Herr_2018_lattice_surgery_mbqc} proposes first to shift part of the cluster state corresponding to one patch, so that the output surface code after measurements corresponds to the input one up to a transversal Hadamard. The rotation is implemented by connecting the output surface code to another cluster and performing local measurements in $X$ or $Z$ basis.

The Hadamard defect we propose here uses a similar principle but in the `hole braiding' framework, where there is only one patch of cluster state for the whole circuit. The defect can be embedded locally within the cluster lattice, creating a fusion surface across which the gate is realized, thereby providing a particularly simple topological implementation.

Promising physical implementations of surface code computation have been proposed and achieved for different architectures (for instance, photonic hardware \cite{Herrera2010Photonic_cluster, Xanadu_package, Bourassa2021blueprintscalable} and superconducting processor \cite{Krinner_2022,Zhao2022_surcae_superconducting,2024GoogleBelowTrheshold} for a non-exhaustive list). Indeed, storing information on surface code patches has been shown to be less resource consuming than holes in the Toric code \cite{fowler2019lowoverheadquantumcomputation}. For now, only relatively small surface codes or cluster states can be tested experimentally. Important scaling up would be necessary to implement a hole-braiding computation on a hardware.


\section{Fault-tolerant computation}
\label{sec:clifford + T gate}
To achieve universality in Quantum computing, one can use the group of logical gates generated by the Clifford group and the $\pi/8$ gate T. In this section, we show that the entire Clifford group can be implemented fault-tolerantly. Precisely, there are measurement patterns and corresponding compatible surfaces that implement any Clifford unitary in the sense of Theorem~\ref{Thm: Fundamental}. Measurement patterns for the identity gate and the $\CNOT$ have been presented in Sec.~\ref{subsec: CSS gates}. In this framework, the computational power is quite limited, not even achieving the whole real Clifford group. To improve this, we propose a new implementation of the Hadamard gate $H$, see Fig.~\ref{fig: Hadamard defect in lattice bis}, which in turn will allow building the phase gate $S$, thus leading to the full Clifford group. Finally, full universality is obtained by adding the $T$ gate, which can be implemented through magic state injection, which will be discussed in Sec.~\ref{Sec: Performance}.

\subsection{Defect in the lattice: the Hadamard gate}
\label{subsec: Hadamard defect}

We will first see that a Hadamard gate on all logical qubits is automatically implemented between two planes separated only by half-odd number of unit cells. Of course, such a global unitary is of no interest, and we would like only some chosen logical qubits to undergo a Hadamard, not all of them. So we would like the lines of some qubits to travel $n+\frac{1}{2}$ unit cells and others only $n$. As we shall see, this can be achieved by introducing a \emph{dislocation in the physical lattice} underlying the cluster state. 

\subsubsection{Primal vs Dual planes: a global Hadamard}

The natural setting of Theorem~\ref{Thm: Fundamental} is where there is an integer number of unit cells between the initial and final planes. We can extend the cluster by one half of a unit cell in the `time' direction, i.e. adding a dual layer after the output plane in Fig.~\ref{fig: Cluste state unit cell}b. By measuring the qubit on the former output plane in the $X$ basis, the arguments of Section~\ref{Sec:Cluster} yield that the former output Toric code is mapped to its dual on the new final plane, see also the foliation principle \cite{Bolt2016_FoliationCSS}. Namely, all the site stabilizers on the black Toric code are transformed into plaquette stabilizers on the red one, and vise-versa. Similarly, site holes implementing primal logical qubits are mapped to plaquette holes, and so the logical operators on the final surface code are mapped to their dual. It follows that the claim of the theorem now holds with $U$ replaced with $ H U $, where $H = \bigotimes_{i=1}^n H_i$. In other words: extending the measurement pattern by half a unit cell corresponds to carrying out one \emph{global} Hadamard gate. In fact, this is exactly what is used in~\cite{Herr_2018_lattice_surgery_mbqc} to implement a logical Hadamard on a foliated surface code.

However, as pointed out above, this is insufficient for our setting where we use only one global cluster. Therefore, our proposal goes beyond that by embedding a lattice defect in the bulk of the cluster. This creates what we shall refer to as a fusion surface that implements a \emph{local} Hadamard gate. It is local in the sense that only those correlation surfaces that cross the fusion surface undergo the gate, but not those that pass around it.

\subsubsection{A line defect}

\begin{figure}[h!]
	\centering
	\includegraphics[width=0.8\textwidth]{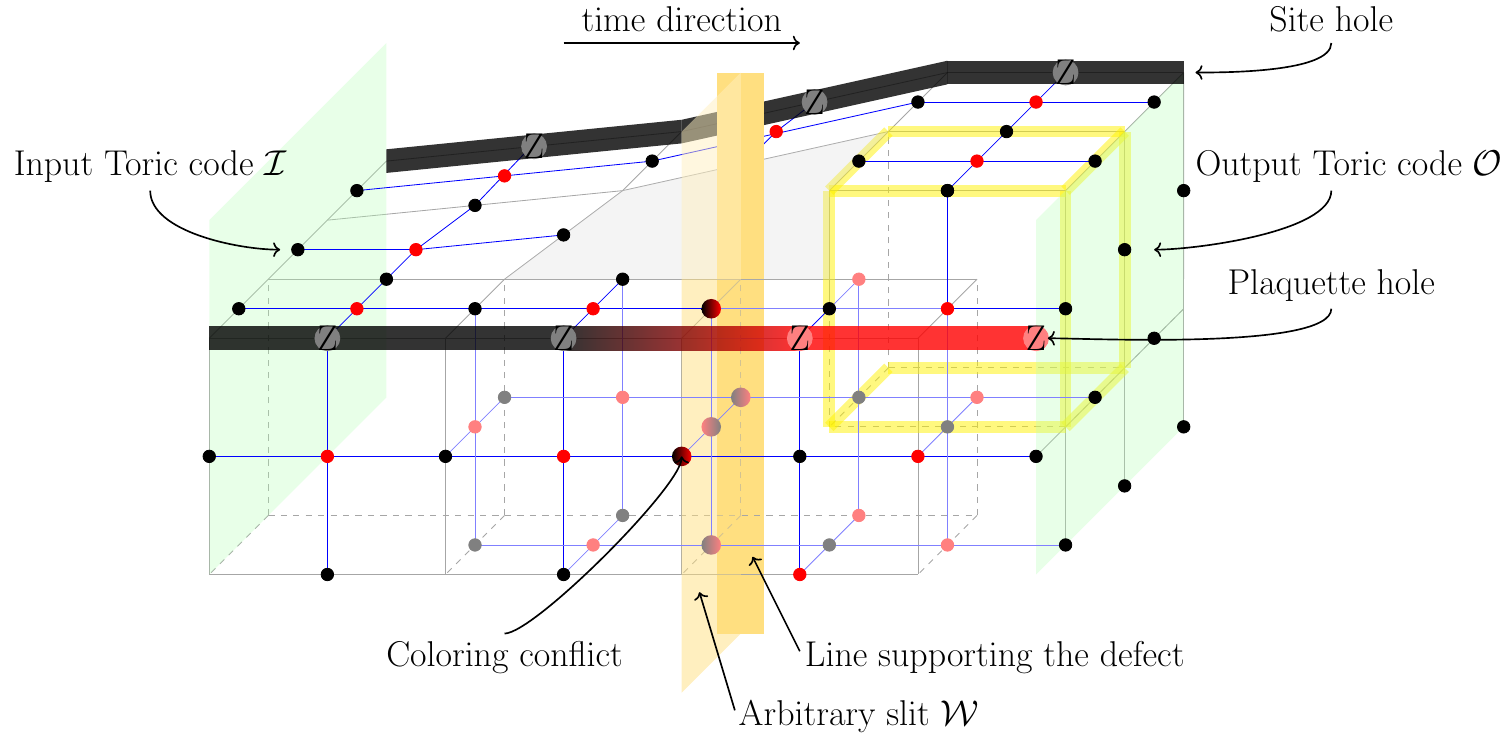}
    \caption{Hadamard defect in lattice at the unit cell level. For clarity, only the qubits on the external faces of the unit cells have been represented, except for 2 unit cells at the front where all qubits are shown. Every cell that is represented by a white cube (potentially distorted) is full: it has a qubit on every face and on every edge. On the output Toric code plane, the red qubits inside the plaquettes that are measured in X are not shown, only the one measure in X and that implements the plaquette hole is displayed. Nearest neighbor connections are represented by blue lines. Some qubits and connections are missing in the middle of the cluster to allow for the dislocation defect.} 
	\label{fig: Hadamard defect in lattice bis}
\end{figure}

Consider the dislocation defect in Fig.~\ref{fig: Hadamard defect in lattice bis}. The lattice is twisted so that the unit cell on the top right (highlighted in yellow) is displaced compared to its normal position. Its left bottom corner now sits on the center of another unit cell. Some of its edges now also correspond to the faces of the other cell. Therefore, trying to bicolor all the qubits from the orange plane in the middle, all the way around the defect and back to the orange plane leads to a coloring conflict.

With this, the lattice of qubits is no longer bipartite, and the duality defined and used above is broken. A connected primal $1$-chain starting on one side of the defect and winding around it once ends up displaced so that it appears as a dual $1$-chain. Equivalently, there is a discrepancy emerging between two connected qubit lines traveling from the initial plane on different sides of the defect: if one of them extends for $n$ unit cells and past the defect the other travels $n\pm\frac{1}{2}$. This is illustrated in Fig.~\ref{fig: Hadamard defect in lattice bis}. Two lines are initially of the same type, both made of Z-measured black edge qubits and accounting for site holes on the input Toric code $\mathcal{I}$ (left green plane), see sick black thick lines in Fig.~\ref{fig: Hadamard defect in lattice bis}. After passing the defect, they end up differently on the right green plane $\mathcal{O}$: The line having passed `behind' the defect still corresponds to black edges and a site hole on $\mathcal{O}$, while the other one accounts now for a plaquette hole, as it sits on Z-measured red faces. In this picture, we assume that all the other surrounding qubits are measured in X except the ones in $\mathcal{O}$.

This defect can be embedded in the 3D lattice, by extending it into a line. It is topological in the sense that the lattice of qubits is locally completely normal. The cluster state can still be built using Eq.~\ref{eq: construction of the cluster state}, and its stabilizers are still defined by $K(a)= X_a \bigotimes_{b\in \partial a} Z_b$, with $\partial a$ the nearest neighbors of $a$. Except for the qubits sitting on the boundary of the defect for which some of their nearest neighbors have been removed, they all still have 4 nearest neighbors and the graph is well defined. The defect can only be detected by winding around it. Furthermore, the input and output Toric code are not geometrically impacted by the defect; in Fig.~\ref{fig: Hadamard defect in lattice bis}, they are made of a square lattice with black qubits on its edges.

\begin{figure}[h!]
	\centering
	\includegraphics[width=0.5\textwidth]{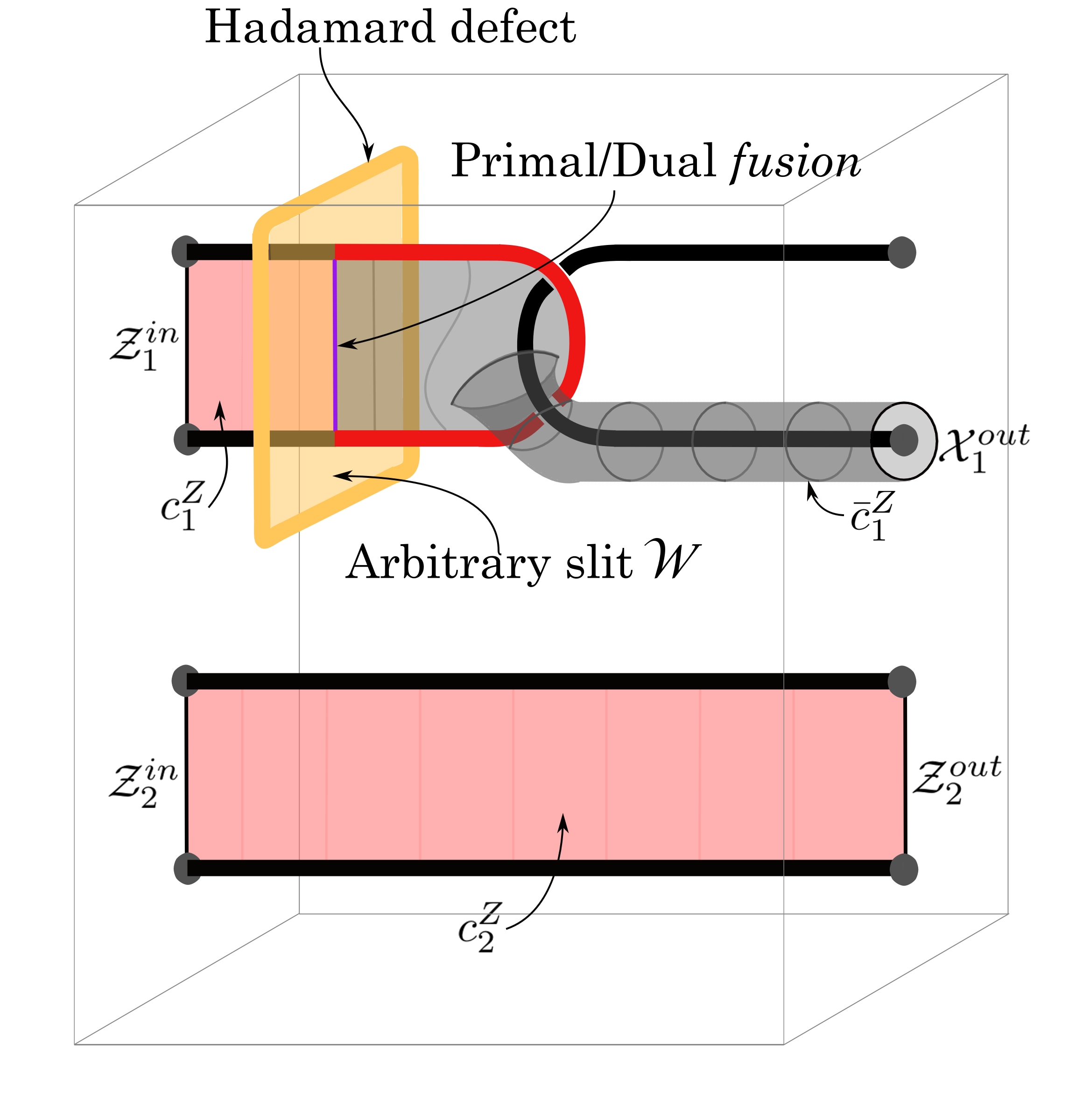}
	\caption{Hadamard defect and arbitrary virtual boundary connecting primal and dual surfaces. Thick black lines correspond to Z measurements on black qubits. Thick red lines to Z measurements on red qubits. }
	\label{fig: Hadamard defect qubits}
\end{figure}

Theorem~\ref{Thm: Fundamental} still applies in this new configuration. We consider a defect line forming a (large) closed loop inside the bulk and we let $\mathcal{W}$ be a surface supported by that loop, see Figure~\ref{fig: Hadamard defect qubits}. In the homological description introduced in Section~\ref{Sec:Cluster}, $\mathcal{W}$ is an internal slit that must be removed from the cell complex. The slit lifts any discrepancy. The duality is completely recovered, with a well-defined final plane $\mathcal{O}$ which we again assume to be a primal surface, namely (\ref{Primal In and Out planes}) holds true. It is important to note that $\mathcal{W}$ is not a domain wall as described in \cite{Bombin_2010,Kitaev_2012} since the lattice shows no defect on $\mathcal{W}$ but only on its boundary. In particular, any surface as described above can be chosen.

The cluster state is defined as before by the graph of nearest-neighbors and so surfaces remain associated with stabilizers, \emph{even across the slit}. From the point of view of homology, however, an `incoming' surface $c\in C_2$ must stop on $\mathcal{W}$ and be connected to an outgoing surface $\bar c\in\overline C_2$ `emerging' from $\mathcal{W}$. These are going to be such that $\partial c\cap \mathcal{W} = \partial \bar c\cap \mathcal{W}$, illustrating again the ambiguity of the nature of the slit (note the slight abuse here: we extend $c,\bar c$ to $\mathcal{W}$). If we introduce the \emph{fusion} operator~$\boxplus_{\mathcal{W}}$ in this situation, which is so that $c \boxplus_{\mathcal{W}} \bar c$ is one surface without boundary on $\mathcal{W}$, the corresponding stabilizer is simply $K(c \boxplus_{\mathcal{W}} \bar c)$ and the surface is compatible with the measurements. Such surface $c^Z_1 \boxplus_{\mathcal{W}} \bar c^Z_1$, is displayed in Fig.~\ref{fig: Hadamard defect qubits}, the fusion happens on the purple line when the two surfaces meet on the slit $\mathcal{W}$. In order to have the same type of logical qubits for all (site holes in the Toric code), we can braid the red Z-measurement lines coming from the slit with black Z-measurement lines, as shown in Fig. \ref{fig: Hadamard defect qubits}. This only performs an identity gate between the plaquette-missing logical qubit after the slit and the site-missing logical qubit on the output plane.   

It remains to observe that
\begin{equation*}
    \logicZ^{\mathrm{in}} \rightarrow \logicX^{\mathrm{out}},\qquad
    \logicX^{\mathrm{in}} \rightarrow \logicZ^{\mathrm{out}},
\end{equation*}
to conclude with Theorem~\ref{Thm: Fundamental} that $U = H_1$ indeed. 

This scheme is fault-tolerant provided the Z-measurement lines remain at a certain distance of the line defect and the circumference of the defect line itself is large enough. Locally the cluster unit cells are preserved and so  error correction can be performed as usual, the Hadamard defect being interpreted merely as an internal cluster boundary.

\subsubsection{Comparison with twists and domain walls}

A parallel can be drawn between the defect in the 3D lattice just discussed and the twist introduced in~\cite{Bombin_2010} on the one hand, or the domain walls of~\cite{Kitaev_2012}, both in the 2D lattice of the Toric code. In all cases, the defect breaks the global duality of the lattice, introducing a map between the electric and magnetic sectors. In~\cite{Bombin_2010,Kitaev_2012}, the analysis is from the point of view of the anyon theory. In general, lattice defects allow for anyon transmutation and impact anyon braiding. In the case of the Toric code, braiding an electric charge around a twist maps it to a magnetic one and vise-versa, which reduces the number of anyonic superselection sectors. The computational implementation of Bombin's lattice defects proposed in~\cite{hastings2015reduced} is different from ours in that it ultimately uses braiding of twists around each other (see also~\cite{Yoder2017surfacecodetwist} for a more complete discussion), while we braid a twist around holes. The hybrid scheme proposed in \cite{Browmn_2017_braidingHolesAndTwists} braids Bombin's twist with holes in a planar setup. However, our defect is microscopically quite different: Rather than being purely spatial, it distorts the space-time lattice. In fact, the shift by one half time step induced by the defect is what realizes the gate, `changing the color' of measurement lines. In the homological picture, this amounts to the transformation of a primal surface into a dual one. Furthermore, locally, the lattice on which is supported the measurement line is untouched, the local stabilizers are preserved. The measurement lines travel only half the time step more than the one that does not braid with the defect (see Fig.~\ref{fig: Hadamard defect in lattice bis}), while in \cite{Browmn_2017_braidingHolesAndTwists} the hole crosses a line where the surrounding stabilizers have to be modified.

\subsection{Rebit encoding and complex S gate}
\label{subsec: Rebit encoding}
In our arsenal, we now have the logical qubit creation and measurement in the X and Z bases \cite{raussendorf_topological_2007}, the $\CNOT$ and Hadamard gates. Thus, the accessible qubit states are, in fact, rebits, namely their density matrix $\rho$ is real
\begin{equation}
	\bra{v}\rho\ket{w} \in \mathbb{R}
\end{equation}
for any $\ket{v}$ and $\ket{w}$ in the computational basis. In order to go further, we use the Rudolph-Grover rebit encoding, see~\cite{rudolph20022}, which uses the simple observation that any complex state
\begin{equation*}
  \left| \psi \right\rangle = \sum_{\mathbf{v} \in \mathbb{Z}^n_2 } r_\mathbf{v} e^{i\theta _\mathbf{v}} \left| \mathbf{v} \right\rangle  
\end{equation*}
of $n$ qubits can be encoded in $n+1$ rebits as follows
\begin{equation}\label{eq: n+1 rebit encoding}
   \left|  \psi^{RE}\right\rangle = \sum_{\mathbf{v} \in \mathbb{Z}^n_2 } r_\mathbf{v} \cos(\theta _\mathbf{v}) \left| \mathbf{v} \right\rangle \otimes \left|R\right\rangle +  r_\mathbf{v} \sin(\theta _\mathbf{v}) \left| \mathbf{v} \right\rangle \otimes \left|I\right\rangle.
\end{equation}
Here we denote $\ket{R}=\ket{0}$ and $\ket{I}=\ket{1}$ the two basis states of the additional rebit. Real gates and X,Z measurements are operations that do not affect the complex structure of the encoded state. Therefore, they can simply be applied without modifying the extra $I/R$ rebit, and can be written schematically:
\begin{equation*}
	\begin{quantikz}
		\lstick{$i$}& \qw  & \meter{X/Z}
	\end{quantikz} \quad \longleftrightarrow
	\begin{quantikz}
		\lstick{$i$}& \qw  & \meter{X/Z} \\
		\lstick{$I/R$} & \qw & \qw & \qw
	\end{quantikz}
\end{equation*}

\begin{equation*}
	\qquad \qquad \qquad\begin{quantikz}
		\lstick{$i$} & \targ{} & \qw &   \\
		\lstick{$j$} & \control{} \vqw{-1}  & \qw & 
	\end{quantikz}  \longleftrightarrow\quad
	\CNOT^{RE} = \begin{quantikz}
		\lstick{$i$} & \targ{} & \qw &   \\
		\lstick{$j$} & \control{} \vqw{-1}  & \qw &   \\
		\lstick{$I/R$} & \qw & \qw &  
	\end{quantikz}
\end{equation*}

Universal Quantum computing can be achieved with a restricted set of gates, namely Paulis, $\CNOT$, $H_{tot}= \bigotimes_{i=1}^{n+1} H_i$, provided the input is in any rebit state, see~\cite{Delfosse_2015}. In the fault-tolerant framework presented above, being able to perform the Hadamard on only one rebit is a considerable advantage (no need for extra ancillas). However, universality is not yet achieved since the initial states at our disposal are restricted to eigenstates of real Pauli operators.

\subsubsection{Encoding of the Y eigenstate}
First of all, the Y eigenstate needs to be encoded: if we denote $\left| \circlearrowleft \right\rangle= \frac{\left| 0 \right\rangle + i \left| 1 \right\rangle}{ \sqrt{2}}$, then its encoded version, following~(\ref{eq: n+1 rebit encoding}), is
\begin{equation*}
		\left| \circlearrowleft^{RE}\right\rangle= \frac{\left| 0 \right\rangle \otimes \left|R\right\rangle + \left| 1 \right\rangle \otimes \left|I\right\rangle}{ \sqrt{2}}
\end{equation*}
Hence, the rebit state $\left| \circlearrowleft^{RE}\right\rangle$ is a Bell state and is accessible fault-tolerantly using the following circuit made of CSS gates only: 
\begin{equation*}
	\begin{quantikz}
		\lstick{$\ket{+}$} & \control{} & \qw   \\
		\lstick{$\ket{R}$} & \targ{} \vqw{-1}  &  \qw
	\end{quantikz} \ket{\circlearrowleft^{RE}}
\end{equation*}

\subsubsection{The complex Clifford gates}
We turn to the implementation of the missing Pauli $Y$ and the $S$ gates on a single rebit encoded qubit. For this, we first note that the multiplication by $i$ amounts to a $\pi/2$-rotation in the complex plane and so $\mathrm{Re}(ie^{i\theta}) = -\sin\theta$ while $\mathrm{Im}(ie^{i\theta}) = \cos\theta$ we conclude with~(\ref{eq: n+1 rebit encoding}) that
\begin{equation*}
    \ket{(i\psi)^{RE}} = 1\otimes XZ\ket{\psi^{RE}}
\end{equation*}
which we shall denote as $i^{RE}= 1\otimes XZ$ (the last tensor factor always refers to the additional $I/R$ qubit). Since $Y = i XZ$, we conclude that
\begin{equation*}
    Y_j^{RE} = X Z \otimes X Z
\end{equation*}
In particular, it is achieved by CSS gates, hence implementable topologically. 

Finally, we turn to the $S$ gate
\begin{equation*}
    S= \begin{pmatrix}
		1 & 0\\
		0 & i
	\end{pmatrix},\qquad
 SXS^\dagger=Y,\, SZS^\dagger=Z.
\end{equation*}
In~\cite{raussendorf_topological_2007}, the implementation of this gate in the framework of topological MBQC required magic state injection, and therefore was not fault-tolerant. In the rebit encoding scheme we propose here, the encoded $S^{RE}$ can be obtained using purely real Clifford gates. With~(\ref{eq: n+1 rebit encoding}) again, we conclude that
\begin{equation*}
    \ket{(S\psi)^{RE}} \equiv \CNOT_{j, I/R} CZ_{j, I/R} \ket{\psi^{RE}}.
\end{equation*}
where the index $j$ refers to the $j^{th}$ rebit. This is a combination of real Clifford gates. Explicitly, we use $CZ_{j, I/R} = H_{I/R}\CNOT_{j, I/R}H_{I/R}$ to obtain the following topological implementation of the $S$ gate
\begin{equation*}
	S^{RE} = 	\begin{quantikz}
		\lstick{$j$} & \qw &  \ctrl{1}  & \qw &   \ctrl{1} & \qw\\
		\lstick{$I/R$} & \gate{H} & \targ{} & \gate{H} & \targ{}&\qw
	\end{quantikz}
\end{equation*}

This representation emphasizes the importance of having a topological implementation of the Hadamard gate. In the above discussions, the rebits can be replaced by encoded rebits on the Toric code. Therefore, we have shown that the complete Clifford group can be implemented topologically, hence in a fault-tolerant fashion, in the framework of MBQC in the cluster state. This is a crucial improvement over previous proposals for topological MBQC on the Toric code, and we will quantify its performance gain in the following Section~\ref{subsec: Overhead clifford}.

\subsection{Overhead for the Clifford gates}
\label{subsec: Overhead clifford}
The operational overhead quantifies the resource cost (number of physical qubits or basic operations in our case) of the computation. Reducing the circuit volume is key due to the practical difficulty of producing large scale quantum circuits. On the other hand, scaling up the size of the cluster exponentially reduces the logical error rate. It is therefore necessary to find an optimal trade-off between cluster state volume and correctability. In this section, we consider for any gate $G$, the ratio between the operational volume and the probability of success of a circuit made of many $G$ gates.

Let $V_G$ be the smallest number of unit cells required to implement the logical gate $G$. Fault-tolerance requires this volume to be scaled up by a factor $\lambda$. Since every elementary cell is built using 24 operations (6 qubit initializations, 12 $CZ$ gates, 6 measurements), the total number of operations required for one $G$ gate is $24 V_{G} \lambda^3$. 

Following the threshold and overhead study in \cite{raussendorf_topological_2007}, the single-gate error $\epsilon_{\mathrm{topo}}(G,\lambda, d)$ is generally dependent on the scaling factor $\lambda$ and the circumference $d$ of the qubit lines. It is given by
\begin{equation}
	\epsilon_{\mathrm{topo}}(G, \lambda, d)= \lambda L_G \left[ e^{-4\kappa(d+1)} +2(d+1) e^{-\kappa(\lambda -d)}\right]
	\label{eq: epsilon topo}
\end{equation}
where the first term corresponds to the probability of a full cycle of errors surrounding one qubit line, while the second term arises from a line of errors emerging from one qubit and ending on another one. $L_G$ is the total length of the Z-measurement lines implementing the gate. The parameter $\kappa$ depends on the physical error. For the purpose of comparison, we shall use below the value $\kappa=0.93$ calculated in~\cite{raussendorf_topological_2007} for a physical error $p=p_{th}/3$, $p_{th}=6.7\times 10^{-3}$.

If a circuit is made of $\Omega$ gates $G$, and we assume that the probabilities of single-gate errors are independent, the success probability of the entire circuit can be approximated by $e^{-\epsilon_{\mathrm{topo}}(G,\lambda, d)\Omega}$. Therefore, the ratio between the total circuit volume and the success rate becomes: 
\begin{equation}
	R_G(\Omega,\lambda,d) =  24 \lambda^3 V_G \Omega e^{\epsilon_{\mathrm{topo}}(G,\lambda, d)\Omega}
	\label{eq: overhead clifford gate}
\end{equation}
Note that $R_G(\Omega,\lambda,d)$ can also be interpreted as the effective operational volume for the circuit. In fact, the quantity $e^{\epsilon_{\mathrm{topo}}(G,\lambda, d)\Omega}$ also approximates the average number of trials needed before success. For circuits on which the final results can be classically verified, a failure would be discarded and the process would then be repeated until success.

\begin{figure}[h!]
	\centering
\begin{subfigure}{0.325\textwidth}
    \includegraphics[width=1\linewidth]{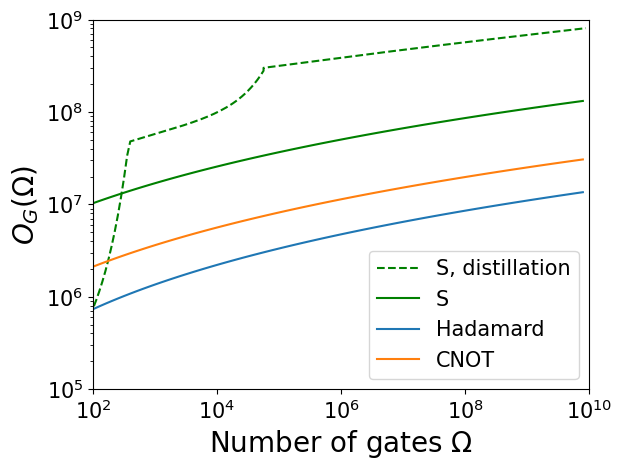}
\end{subfigure}
\begin{subfigure}{0.325\textwidth}
    \includegraphics[width=1\linewidth]{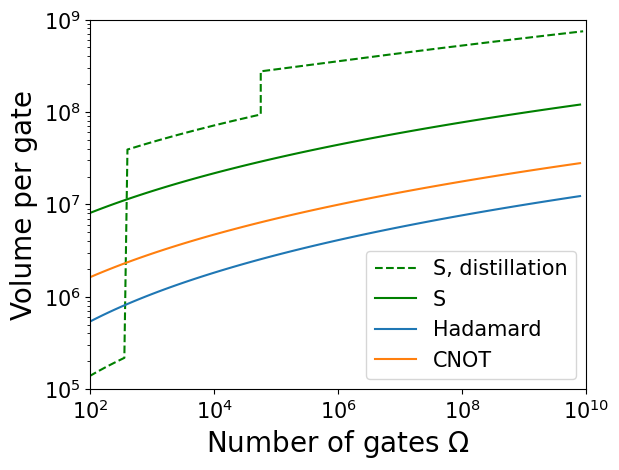}
\end{subfigure}
\begin{subfigure}{0.325\textwidth}
    \includegraphics[width=1\linewidth]{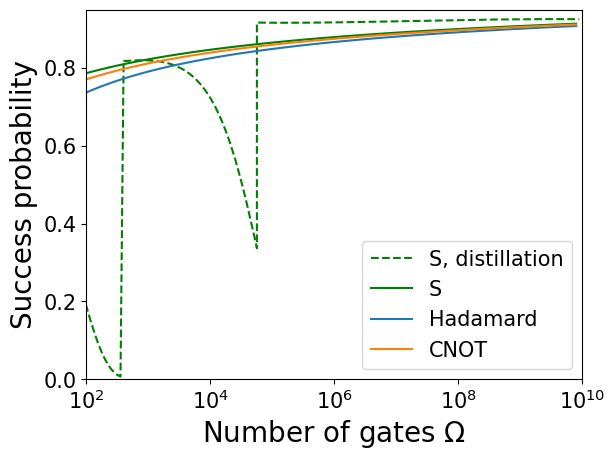}
\end{subfigure}
\caption{The overhead $O_G(\Omega)$ for the $H$, $\CNOT$ and $S$ gates. From the optimized ratio $O_G(\Omega)$ (left), we compute the operational volume per gate $24 V_G \lambda_{opti}^{3} $ (middle), and the success probability of the entire circuit $e^{\epsilon_{topo}(G,\lambda_{opti}, d_{opti})\Omega}$ (right). The comparison is particularly striking for the $S$ gate: The jumps in the dotted line correspond to additional rounds of distillation, and they are absent in the topological implementation (the solid green line) proposed in this work.}
    \label{fig: Overhead Cnot and S}
\end{figure}

For the topological gates, $R_G(\Omega,\lambda,d)$ can be easily computed and optimized over the parameters $\lambda,d$. In Figure~\ref{fig: Overhead Cnot and S}, we show the overhead per gate
\begin{equation}\label{eq:Overhead per gate}
    {O_G(\Omega) = \Omega^{-1}\min\left\{R_G(\Omega,\lambda,d):\lambda,d\geq 1\right\}},
\end{equation}
as well as the the volume per gate and the success probability for the optimal $(\lambda_{opti},~d_{opti}$), as a function of the number of gates $\Omega$ for the topological Hadamard, $\CNOT$ and $S$ gates. In the $S$ gate case, we can immediately compare with the previous implementation: Thanks to the topological Hadamard and the rebit encoding, the operational volume associated with the $S$ gate is reduced by one full order of magnitude for large $\Omega$, while the success rate reaches 90\% around $\Omega=10^8$ versus 92\% for the magic state distillation method. In general, the optimization process is such that, for higher $\Omega$, maximizing the success rate will be preferred over minimal volume, due to the exponential dependency in $\Omega$. The volume and length of the gates used in the numerics are listed in Table~\ref{table: gates volume and length}.

We conclude this section by noting that a similar study has been carried in Ref.~\cite{Lee_MBQC_2Dcolorcode} but with the 2D color code instead of the Toric code as the base for the 3D cluster. The self-dual structure of the color code allows for a more direct implementation of the Hadamard and $S$ gates. Indeed, in that framework, the correlation surfaces are more complex and can be compatible with measurements in the complex $Y$ Pauli basis. In this paper, the overhead is defined as the number of physical qubits $n$ or $CZ$ gates $N_{CZ}$ per logical qubit on the 2D base code, as a function of the code distance $d$. This amounts to finding the optimal arrangements for a maximum of logical qubit holes to fit in a lattice of a given size while preserving the logical distance $d$. The comparison between the Toric code based MBQC with the color code one is made for different color code lattices: the ratio $n/k$ is about $6.6d^2$ for the Toric code versus $3.9d^2$ and $3.7d^2$ for 4-8-8 and 6-6-6 color codes respectively. No calculation of the logical gate overhead on the whole 3D cluster has been conducted in the mentioned paper, however it would be of interest to compare the performance of both cluster states.

\section{Universal Quantum Computation: the $T$ gate}\label{Sec: Performance}

In this final section, we complete the Clifford group with the implementation of the $T$ gate and thus achieving universal MBQC. Compared to the gates presented above, the fault-tolerant version of the $T$ gate remains the most expensive one due to the necessary distillation. Nonetheless, we show that a simple geometric optimization using equivalence transformations of the topological circuitry yields a significant advantage.

\subsection{$T$ gate by magic state injection} \label{subsec: universality T gate}
Universal Quantum computing requires the Clifford group to be extended by the rotation gate $ T= \sqrt S = e^{\frac{i\pi Z}{4}}$. This gate cannot be implemented fault-tolerantly in MBQC on 3D cluster states. Nonetheless, fault-tolerance can be achieved by magic state distillation -- details are left for the appendix. Given an ancilla $ \ket{A}= \frac{\ket{0} + e^{i\pi/4} \ket{1}}{\sqrt{2}}$, the $T$ gate can be implemented as follows:
\begin{equation}
	\begin{quantikz}
		\lstick{$\ket{\psi}$} & \targ{} & \qw & \meter{Z}  \vqw{1}\\
		\lstick{$\ket{A}$} & \control{} \vqw{-1} & \qw  & \gate{S} &\rstick{$T\ket{\psi}$} \qw
	\end{quantikz}
	\label{eq: T gate circuit}
\end{equation}
The $S$ gate in the circuit is conditional on the result of the $Z$ measurement and it must carried out on average every second time. A topological $\ket{A}$ state is obtained by measuring just one qubit of a stabilizer on the initial plane in the $X-Y$ plane\cite{raussendorf_topological_2007, raussendorf_fault-tolerant_2006}, which is not a fault-tolerant. In order to recover a high quality topological $\ket{A}$ state, one must resort to magic state distillation, which requires itself 15 ancilla qubits encoded in the Reed-Muller code \cite{steane1996quantumreedmullercodes, knill1996thresholdaccuracyquantumcomputation, Reed_Muller_website}. Repeating the procedure recursively suppresses the error on $\ket{A}$ exponentially, see~\cite{Bravi_Kitaev_magic_distillation} and the appendix, and the state can be used as in~(\ref{eq: T gate circuit}).

The above has all been described in the qubit setting and must now be adapted to our newly proposed rebit setting. The $\ket{A}$ state needs to be encoded in rebits: $$\ket{A^{RE}}= \frac{1}{\sqrt{2}}\left( \ket{0}\otimes\ket{R} + \ket{1}\otimes \frac{1}{\sqrt{2}} (\ket{R} +\ket{I})  \right).$$ This state can be obtained using the (complex) magic $\ket{A}^*$ state, $S$ and $T$ gates by:
\begin{equation}\label{A_RE}
 \begin{quantikz}
		\lstick{$\ket{+}$}&  \ctrl{1} & \qw  & \qw &   \qw & \qw\\
		\lstick{$\ket{A}^*$} & \targ{} & \gate{T}& \gate{H} & \gate{S}&\qw
	\end{quantikz}  \ket{A^{RE}}
\end{equation}
thereby involving magic state distillation, as well as the new topological Hadamard. Injecting the resulting $ \ket{A^{RE}}$ on the left of~(\ref{eq: T gate circuit}) yields the rebit-encoded $T$ gate.

\subsection{Overhead for the $T$ gate}

Temporarily putting aside the rebit encoding, the $T$ gate requires $\ket{A}$ and $\ket{Y}$ states distillation which is resource intensive. Each round of distillation reduces the noise of the ancilla qubit but drastically increases the number of necessary operations. We summarize this in~(\ref{eq: recursion relation},\ref{eq: overhead T gate magic Y}), which follows~\cite{raussendorf_topological_2007} and is detailed in the appendix. If $\epsilon_{l}^A$ is the probability of error on the ancilla after the $l^{th}$ round of distillation and $O^A_l$ is the overhead of the $l^{th}$ round of distillation ($l=0$ corresponds to no distillation at all), then

\begin{equation}
	\begin{split}
 \epsilon_{l}^A&= 35( \epsilon_{l-1}^A)^3 + \epsilon_{\mathrm{topo}}(A, \lambda_{l-1}, d_{l-1})\\
		O^A_l&= \frac{1}{1- 15 \epsilon_{l-1}^A - \epsilon_{\mathrm{topo}(L_{A},} \lambda_{l-1}, d_{l-1})} (15O^A_{l-1} + \frac{1705}{512}O^Y_{l-1} +  24V_{A} \lambda_{l-1}^3)
	\end{split}
	\label{eq: recursion relation}
\end{equation}
where $\epsilon_{\mathrm{topo}}(A, \lambda_{l-1}, d_{l-1})$ is the failure rate of the topological Reed-Muller circuit. For the numerics, we shall use $\epsilon_{0}^A= 0.0134 =6p=2p_{th}$ following~\cite{raussendorf_topological_2007}.

Let $l_\mathrm{max}$ denote the final round of distillation. The overhead for $T$ is computed as in~(\ref{eq: overhead clifford gate}), taking into account the distillations of $\ket A$ and $\ket Y$ as well as the topological circuitry, yielding

\begin{equation}
	O_T(\Omega)=\Omega \left(O^A_{l_{\mathrm{max}}} +(1/2)O^Y_{l_\mathrm{max}} + 36 (\lambda_{l_\mathrm{max}})^3V_T\right)\mathrm{exp}\left(\left(\epsilon_{l_\mathrm{max}}^A + \epsilon_{l_\mathrm{max}}^Y + \epsilon_{\mathrm{topo}}(L_T, \lambda_{l_\mathrm{max}}, d_{l_\mathrm{max}})\right)\Omega\right)
	\label{eq: overhead T gate magic Y}
\end{equation}

\begin{figure}[h!]
		\centering
\begin{subfigure}{0.325\textwidth}
    \includegraphics[width=1\linewidth]{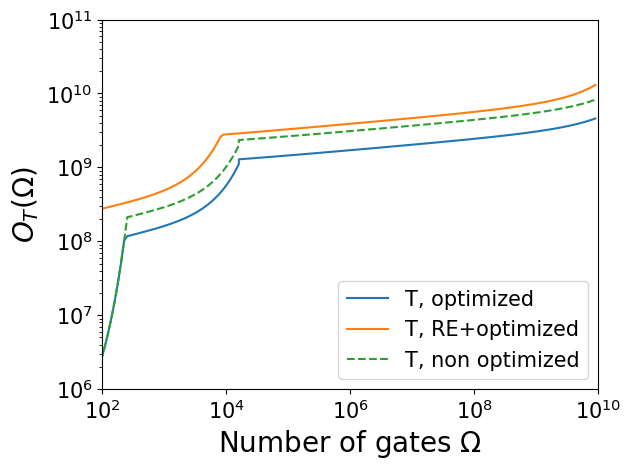}
\end{subfigure}
\begin{subfigure}{0.325\textwidth}
    \includegraphics[width=1\linewidth]{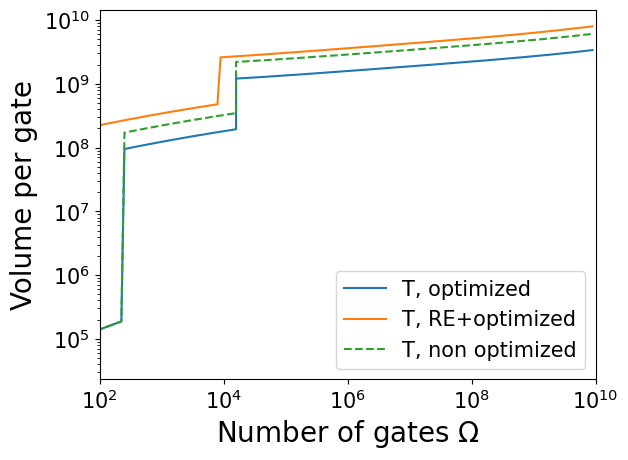}
\end{subfigure}
\begin{subfigure}{0.325\textwidth}
    \includegraphics[width=1\linewidth]{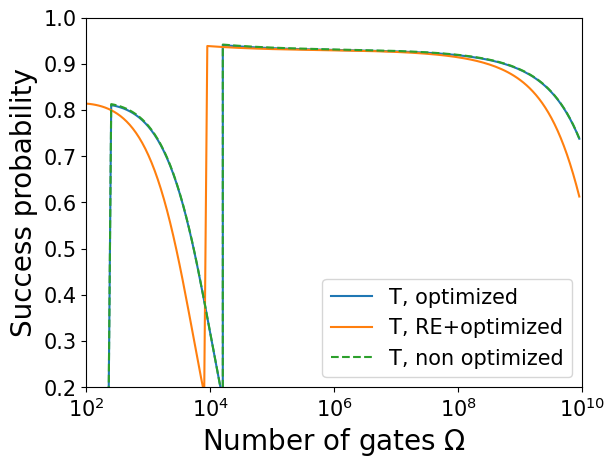}
\end{subfigure}
	\caption{Overhead for the $T$ gate using two different volumes for the $\ket{A}$ state distillation (see Table \ref{table: gates volume and length}). The overhead is also calculated for the $T$ gate in the rebit encoding. Left: optimized ratio $O_T(\Omega)$, middle: volume per gate, right: success probability.  } 
	\label{fig: Overhead T}
\end{figure}

We refer to Figure~\ref{fig: Overhead T} for the resulting numerics. Not surprisingly, because distillation is still required, the improvement over past implementations is slightly less spectacular, but the volume per gate is reduced by one half of an order of magnitude nonetheless. The main contribution to the overhead~(\ref{eq: recursion relation}) is the distillation, and in it the leading term is $\lambda_0 V_A$. The `optimized' curve in the figure is obtained after the fundamental volume involved in the topological implementation of the Reed-Muller code was reduced by more than $40\%$ by a mixture of topological and geometric optimization, see again Table~\ref{table: gates volume and length}. We shall discuss this process in more details in Section~\ref{subsec: circuit volume optimization}.

For our proposal which uses a rebit encoding, the $T$ gate uses one extra magic $\ket{A}^*$ state and more topological circuitry, which increases the overhead. The jumps corresponding to another round of distillation tend to happen for lower $\Omega$ as well. All in all, this $T$ gate volume and success probability (orange curve in Figure~\ref{fig: Overhead T}) is roughly similar to the previous implementation (green-dotted curve). 

\begin{table}[h!]
\centering
\begin{tabular}{||l|l|l||} 
 \hline
 Gate & Volume & Length  \\ 
 \hline\hline
 \CNOT{} (compacted) & $V_{\CNOT}=12$ & $L_{\CNOT}=22$  \\ 
 \hline
 Hadamard & $V_H=6$ & $L_H=6$ \\
 \hline
 S (rebit encoding) & $V_S=48$ & $L_S=48$ \\
 \hline
 A distillation & $V_A=336$ & $L_A=362$\\
 \hline
 A distillation (optimized) & $V_A=192$&$L_A=288$\\
 \hline
 Y distillation & $V_Y=120$ & $L_Y=120$ \\
 \hline
 Y distillation (optimized) & $V_Y=70$ & $L_Y=105$ \\
 \hline
 CSS circuit for T & $V_{T}=2$ & $L_{T}=3$ \\ 
 \hline
 CSS circuit for T (RE) & $V_{T,RE}=89$ & $L_{T,RE}=89$ \\ 
 \hline
\end{tabular}
\caption{Volume and length of different gates and circuits used in the overhead equations. Except from the Hadamard,the S gate and the optimized circuits, those value are the same as in Ref. \cite{raussendorf_topological_2007}.}
\label{table: gates volume and length}
\end{table}

\section{Circuit volume optimization}
\label{subsec: circuit volume optimization}

We finally turn to the circuit optimization problem. By this, we mean the question of the reduction of the fundamental volume $V_G$ for a gate $G$. The fundamental theorem does not claim uniqueness of the measurement pattern for a given $G$. Beyond the topological invariance --- homological surfaces yield the same gate --- which is intrinsic to the framework, there are other equivalent measurement patterns that are not homological. In this section, we first formalize the notion of local equivalence between measurement patterns. We then introduce a new set of non-topological transformations that have not yet been proposed in the literature to our knowledge. Finally, we apply these transformations to the Reed-Muller code and quantify the volume reduction thus achieved. 

Before delving into further details, we would like to review previous works which achieve volume optimization for braiding computation in the Toric code \cite{paetznick2013quantumcircuitoptimizationtopological, fowler2013bridgeloweroverheadquantum, Hanks2020CompressionWithZXCalculus}. Note that these works aim at optimizing the space-time volume of a circuit. The physical setup is a Toric code with hole defects that move in time, and this motion is represented as lines in a (2+1)D diagram. These lines have a direct translation into the 3D cluster state: they correspond to $Z$-measurement lines. Therefore, the methods presented in \cite{paetznick2013quantumcircuitoptimizationtopological, fowler2013bridgeloweroverheadquantum, Hanks2020CompressionWithZXCalculus} are totally applicable to MBQC. First, \cite{paetznick2013quantumcircuitoptimizationtopological} proposes two algorithms for compacting circuits topologically (lines can be deformed arbitrarily as long as they are not cut or they do not merge). Beyond topological transformations, \cite{fowler2013bridgeloweroverheadquantum} uses `bridges' transformation to optimize $\CNOT$ based circuits. The authors show that introducing a bridge between two $\CNOT$, i.e. drawing a red line between two loops such as in Fig.~\ref{fig: CNot}, does not change the computation. They apply this transformation to distillation circuits (Steane code and Reed-Muller code) and perform further topological optimization as in \cite{paetznick2013quantumcircuitoptimizationtopological}. For the Reed-Muller code, they obtain a final volume of 192, a volume 1 in their frame work would correspond to one unit cell for the 3D cluster. More recently, \cite{Hanks2020CompressionWithZXCalculus} translates the 3D topological representation into circuits in $ZX$ calculus \cite{Coeke_2008_ZX1, Coecke_2011ZX2}. The authors then rewrite the circuits using $ZX$-calculus rules and translate them back into the 3D representation. For the Reed-Muller code, they achieve a volume reduction of 65\%, or a final volume of 125. Note that the works presented above optimize preexisting diagrams corresponding to some given circuits, while \cite{Paler_2016_circuits_to_mbqc} proposes algorithmic tools to translate any quantum circuit into a topological MBQC pattern.

\subsection{Equivalence between two measurement patterns}

We start with a notion of \emph{local equivalence} of measurement patterns. Importantly, it refers only to the \emph{lines of measurements} and not to the surfaces (and hence qubits) associated with them: It is an equivalence that holds for any surface compatible with the lines under consideration. 

\begin{figure}[h!]
    \centering
    \begin{subfigure}[r]{0.63\textwidth}
    \centering
        \includegraphics[width=1\linewidth]{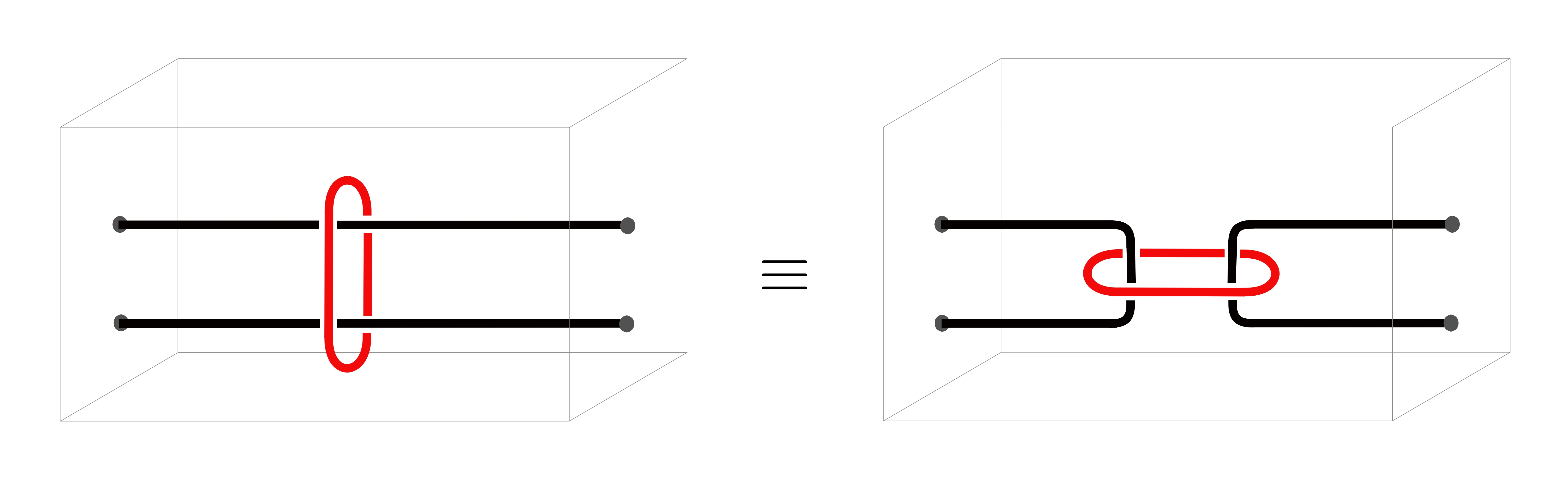}
        \caption{Equivalence of a dual loop around two primal lines.}
    \end{subfigure}
    \\
    \begin{subfigure}[l]{0.63\textwidth}
    \centering
        \includegraphics[width=1\linewidth]{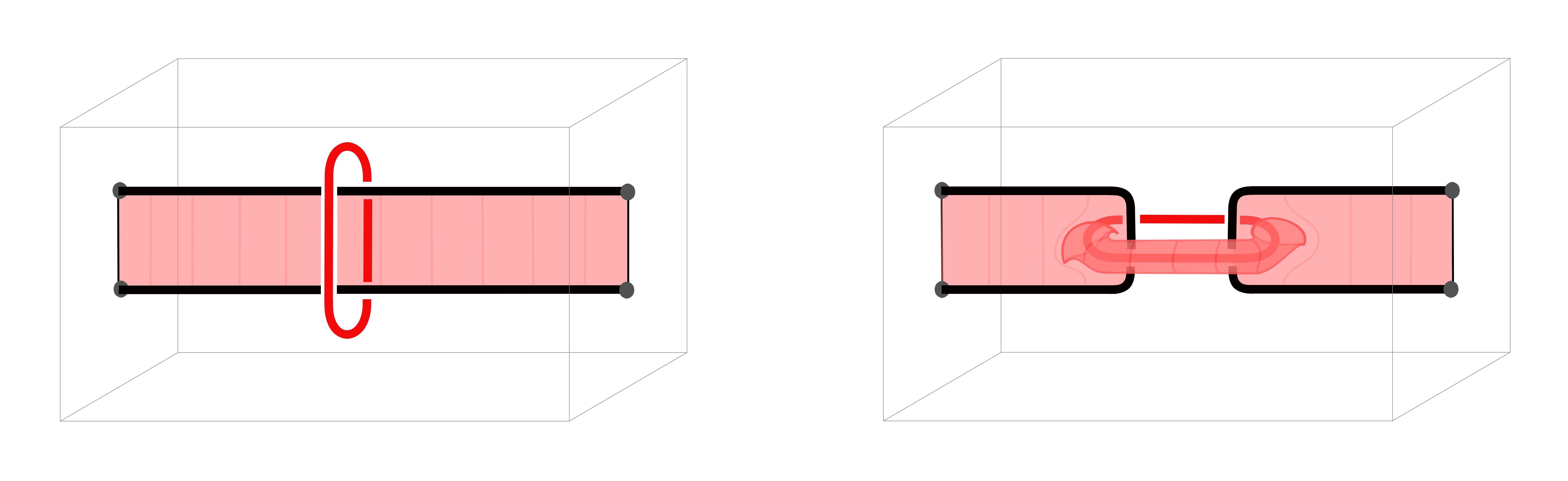}\\
        \includegraphics[width=1\linewidth]{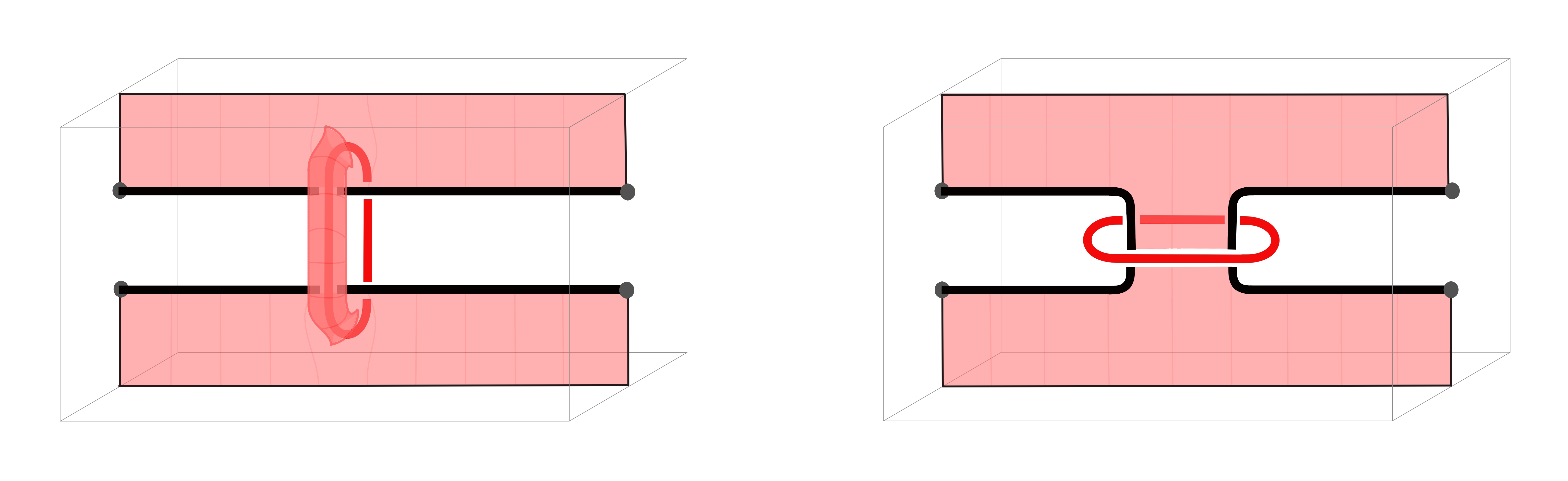}\\
        \includegraphics[width=1\linewidth]{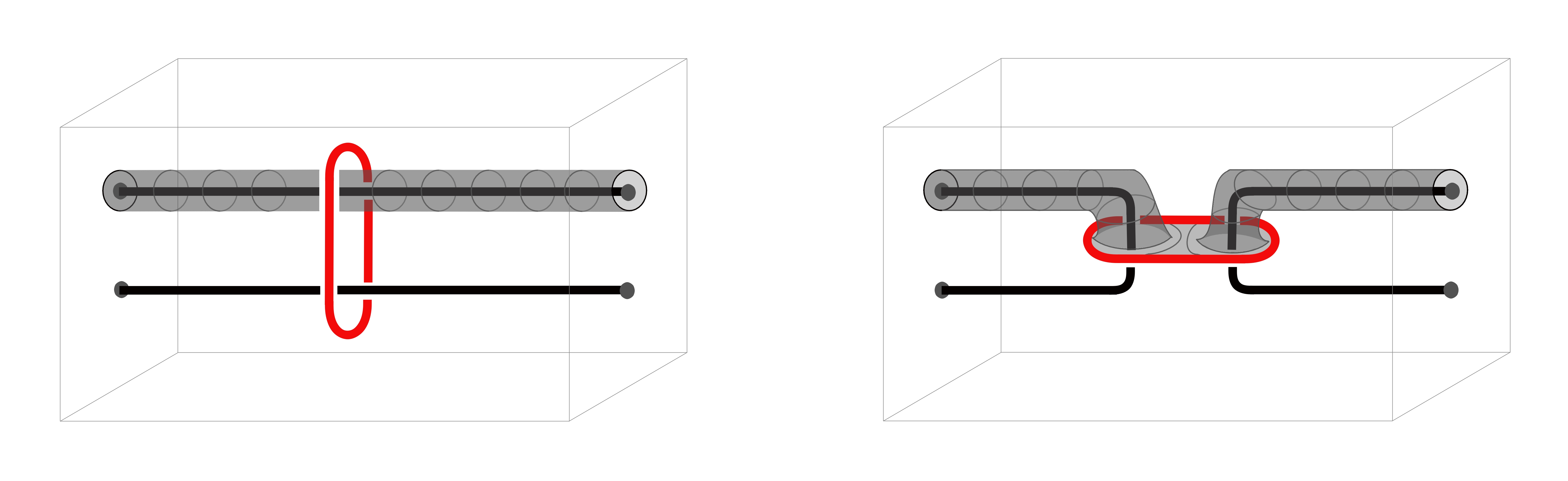}\\
        \caption{Examples of corresponding surfaces in the two equivalent measurement patterns. }
    \end{subfigure}
    \caption{Two equivalent measurement patterns and examples of surfaces that connect the same operators on the boundary of the cube.}
    \label{fig:equivalence Example}
\end{figure}

For simplicity, we consider a cube $B$ of the cluster: by topological invariance, any local patch of measurements can be isolated within a cube. Its surface contains a number of fixed marked (primal or dual) qubit holes that we denote $\{b_1,\ldots,b_m\}\cup \{r_1,\ldots,r_n\}$. Two sets of Z-measurement lines are given by two elements $(c_1,\bar c_1)$ and $(c_1',\bar c_1')$ of $C_1\times\overline{C_1}$. We consider their equivalence under the condition that the marked points on the boundary are fixed, namely
\begin{equation*}
    \partial c_1 = \partial c_1' = \{b_1,\ldots,b_m\}, \qquad 
    \partial \bar c_1 = \partial \bar c_1' = \{r_1,\ldots,r_n\}.
\end{equation*}
Then $(c_1,\bar c_1)$ and $(c_1',\bar c_1')$ are equivalent if, for any surface $c_2$, respectively $\bar c_2$, that is compatible with $(c_1,\bar c_1)$, there is surface $c_2'$,  respectively $\bar c_2'$, that is compatible with $(c_1',\bar c_1')$ and such that the intersections of $c_2$ and $c_2'$, respectively $\bar c_2$ and $\bar c_2'$, with the boundary of the cube $B$ are equal. An example of two equivalent patterns that are topologically inequivalent is given in Figure~\ref{fig:equivalence Example}.

\subsection{Loop equivalence}

\begin{figure}[h!]
    \centering
    \includegraphics[width=0.63\linewidth]{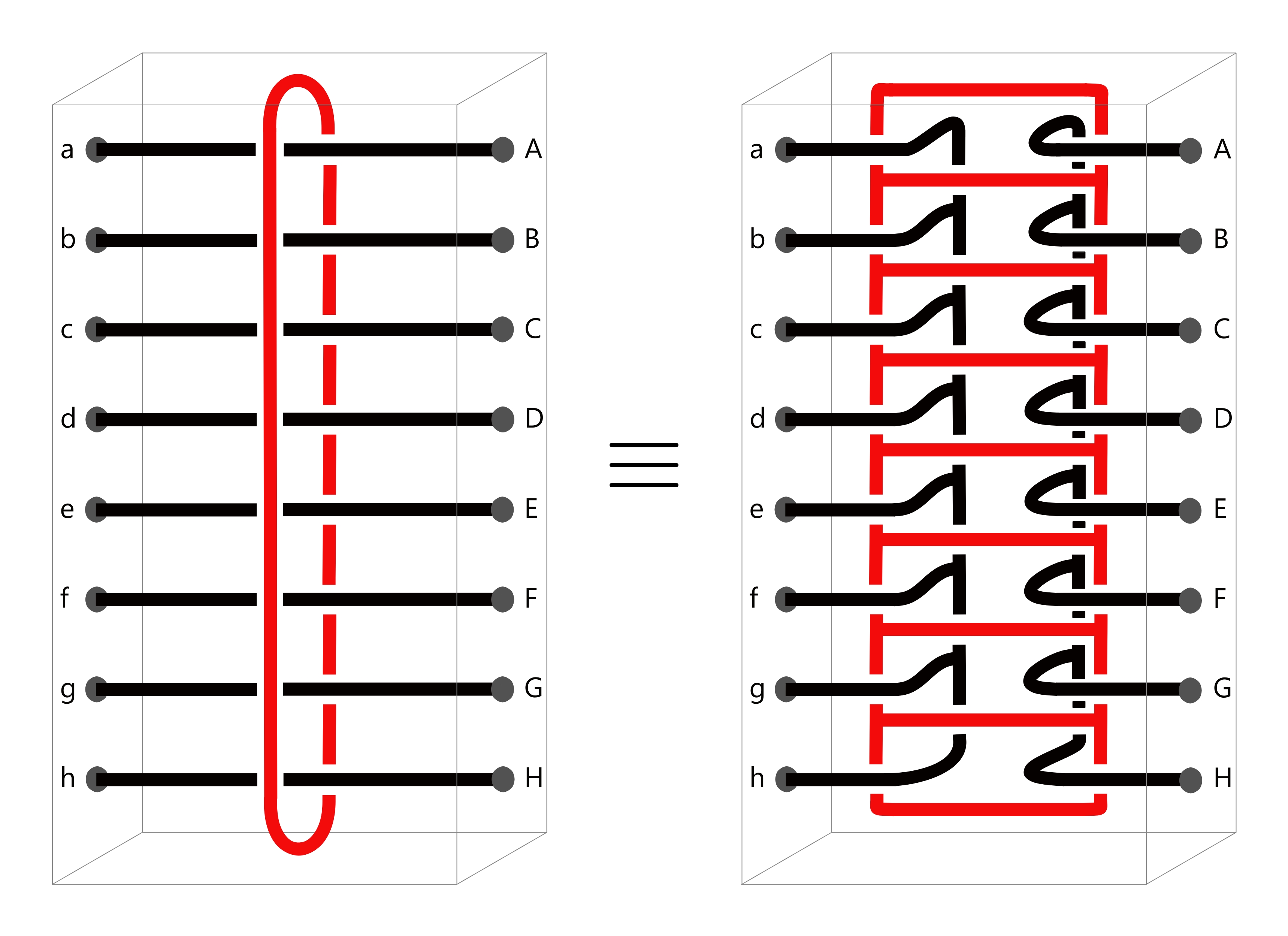}
    \caption{An equivalent pattern for a dual loop around 8 primal lines.}
    \label{fig:Big Equivalence}
\end{figure}

The example in Figure~\ref{fig:equivalence Example} is in fact most relevant as the pattern arises in the $\CNOT$ gate. In fact, it has an immediate generalization to a loop surrounding more than two lines: This will be relevant when we consider the Reed-Muller code, which contains five loops each linking eight defect lines, see Figure~\ref{fig: circuit distillation} in the appendix. The equivalence in this general case is displayed in Figure~\ref{fig:Big Equivalence}.

\subsection{Compacting the magic state distillation circuit}

\begin{figure}[h!]
	\centering
	\includegraphics[width=0.6\textwidth]{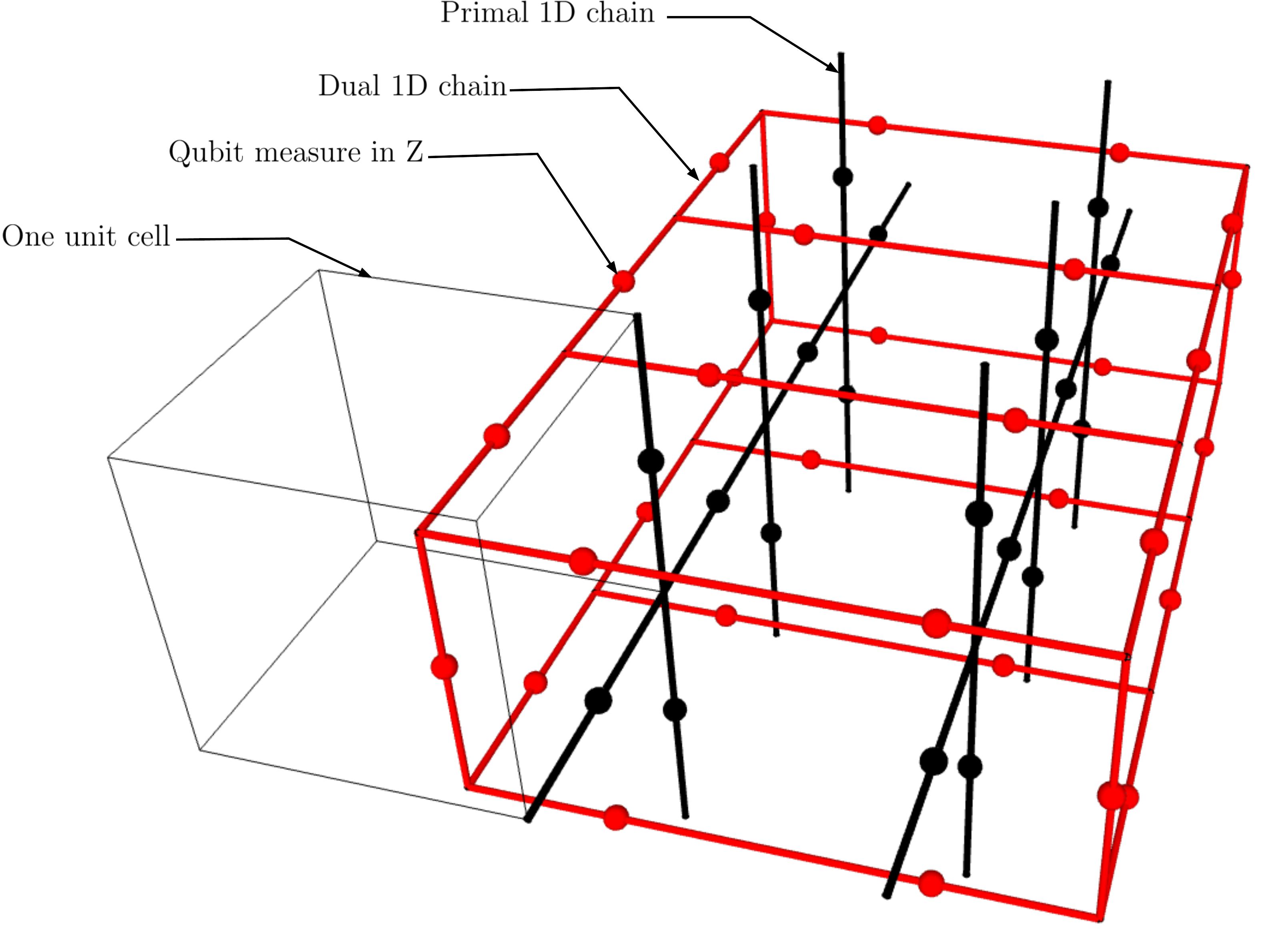}
    \caption{Folding the panel on the right hand side of Figure~\ref{fig:Big Equivalence}. The figure is at the unit cell scale, a primal unit cell is represented by a cube on the front left corner of the box, while the black and red dots represent the black and red qubits, respectively.}
    \label{fig: 8 lines in a rectangle cube}
\end{figure}

We now use the remaining topological equivalence to further compact the Reed-Muller circuit. For this, we again consider Figure~\ref{fig:Big Equivalence}. The red pattern forms a `panel' divided into eight compartments. Folding along the horizontal red line, the panel can be wrapped on itself to form a rectangular cuboid. The result is displayed in Figure~\ref{fig: 8 lines in a rectangle cube}. Not only is this very compact, we also immediately see that the volume is 16 unit cells that are shared with the other boxes. The Reed-Muller code requires 5 of such boxes to be connected appropriately. The geometry proposed in Figure~\ref{fig: circuit distillation}(b) does so in a compact way, resulting in a fundamental volume $V_A$ for the Reed-Muller encoding of 192 unit cells. This is to be compared with the `naive' circuit of Figure~\ref{fig: circuit distillation}(a) in the appendix, which requires 336 unit cells. These are the values used in Table~\ref{table: gates volume and length}. Our optimization compares with the bridge transformations \cite{fowler2013bridgeloweroverheadquantum}, but does not perform as well as \cite{Hanks2020CompressionWithZXCalculus} which achieves a final volume of 125 for the Reed-Muller encoding. The `loop-to-box' transformations we present here is a new method that could be combined with previous ones to further improve MBQC in the 3D cluster state. 

\subsection{Algorithmic Circuit Verification}

Verifying the equivalence of two measurement patterns by hand is a cumbersome process, as this requires checking every possible connection between non-trivial loops on the boundary.
A numerical assistant for this task is glaringly necessary. Problematically, the number of surfaces to check will grow exponentially with the number of Z-measurement lines ending at the boundary, again quickly rendering the problem intractable.
However, the goal of these local equivalence transformations is to reduce the volume of a global logical circuit. It is thereby both sufficient and more efficient to verify that the global circuit is generating the correct logical operator \emph{posterior} to applying all local transformations.
Predicated on this, we have built a program that takes as an input a measurement pattern of a CSS circuit, as well as a set of logical operators on the boundary, and returns whether respective compatible surfaces connecting these logical operators exist.
The specifics of this algorithm are detailed in the Appendix, and its code is available under~\cite{Schwiering2025}.

Every novel CSS circuit in this paper has been verified numerically using that algorithm. Furthermore, to ensure its correctness with a reasonable degree of certainty, it was previously tested with (1) already established circuits with well-known results,\footnote{Such as the familiar \CNOT{} gate visualized in Fig.~\ref{fig: CNot}, which indeed possesses all compatible surfaces given by Eq.~\ref{eq: surfaces CNOT}.} as well as with (2) deliberately wrong constructions to reject the possibility of continuing false positives. As a result, we can argue with high confidence that all novel circuit decompositions presented here are correct and significantly compactified variants of previous iterations. Possible research directions arising from this numerical verification are discussed in the following.

\section{Conclusion and Discussion}

In this paper, we studied the intrinsic relation between a measurement pattern inside the RHG 3D cluster state and the implementation of a gate in a Toric code. We proposed a general theorem using the correlation surfaces supported by the measurement pattern and relating them to the logical real Pauli operators in the Toric code. Thus, we showed the possibility of achieving all the gates in the real Clifford group fault-tolerantly. With that in mind, we introduced a local dislocation defect in the 3D cluster lattice, still compatible with the above theorem, that allows for an intrinsically fault-tolerant implementation of the Hadamard gate. An encoded version of the $S$ gate was then obtained through the Rudolph-Grover encoding, completing the full Clifford group fault-tolerantly. We quantified the performance gains of such structural improvements, reducing the $S$ gate overhead by one order of magnitude, as compared with previous proposals involving magic state distillation.

We showed also significant gains in the $T$ gate overhead. To achieve universal MBQC in 3D cluster states, we still needed to resort to magic state distillation which is resource consuming. By optimizing the geometry of the distillation topological circuit, we managed to reduce the $T$ gate overhead by about half of an order of magnitude. 

Finally, in order to verify the relevance of the optimized circuit, we built a program that checks that a given measurement pattern in the cluster state is indeed achieving the desired logical operation. 

The following questions remain open:

\begin{itemize}
\item The compact realization of the magic state distillation circuit displayed in Figs.~11 and 13 (b) was found by hand. In the continuity of the verifier program, can the search for more compact topological sub-circuits be numerically automated?

\item We have introduced a new defect into the RHG lattice, of disclination type, that brings new topological computational power to 3D cluster states. Following the recent classification of defects in the Toric code  and the color code \cite{Bombin_2010, Kitaev_2012, Barkeshli_TwistDefct2013, Bridgeman_TwistTensorNetwork, Kesselring_2018, kesselring_anyon_2022, Wang_TwistClassification2024}, can the defects of the RHG lattice be classified? What are their computational powers?

\item Moreover, the lattice structure is fundamentally changed by the introduction of this defect. It breaks down the lattice duality, and therefore the notion of chain complex too. In this paper, we deal with it physically by introducing a virtual cut where the lattice ends. Does a more general description of this phenomenon exist at a higher mathematical level? 

\item Finally, the topological framework studied here deals with correlation surfaces only compatible with X and Z measurements, restricting the possibilities of Quantum Computation. In Ref.~\cite{Lee_MBQC_2Dcolorcode}, the foliation of the 2D color code results in correlation surfaces compatible with $Y$ measurement, thus achieving the full Clifford group too. In the light of recent studies on 3D color codes \cite{Davydova2024quantumcomputation, song2024magicboundaries3dcolor}, could a higher dimensional cluster state be built where correlation surfaces would be compatible with non-Pauli measurements? 
\end{itemize}

\section*{Acknowledgments}
We thank \begin{otherlanguage}{lithuanian} Ugnė Liaubaitė,
\end{otherlanguage} Ryohei Weil, and Ruben Campos Delgado for their more than helpful discussions. 

\noindent SB acknowledges the support of NSERC of Canada.

\noindent RR is funded by the Humboldt Foundation.

\noindent This work was supported by NSERC and the European Commission under the Grant \emph{Foundations of Quantum Computational Advantage}.

\bibliographystyle{quantum}
\bibliography{biblio}

\appendix
\section{T gate and magic state distillation}
\subsection{T gate by magic state injection}

The rotation gate $T= e^{\frac{i\pi Z}{4}}$ cannot be implemented topologically in our framework. To circumvent this issue, one can consider the circuit presented in~(\ref{eq: T gate circuit annex}) using the ancilla $ \ket{A}= \frac{\ket{0} + e^{i\pi/4} \ket{1}}{\sqrt{2}}$. The resulting state depends on the $Z$ measurement outcomes, and a $S$ gate needs to be subsequently applied every second times, in average. 

\begin{equation}
	\begin{quantikz}
		\lstick{$\ket{\psi}$} & \targ{} & \qw & \meter{Z}  \vqw{1}\\
		\lstick{$\ket{A}$} & \control{} \vqw{-1} & \qw  & \gate{S} &\rstick{$T\ket{\psi}$} \qw
	\end{quantikz}
	\label{eq: T gate circuit annex}
\end{equation}

\begin{figure}[h!]
	\centering
	\includegraphics[width=0.5\textwidth]{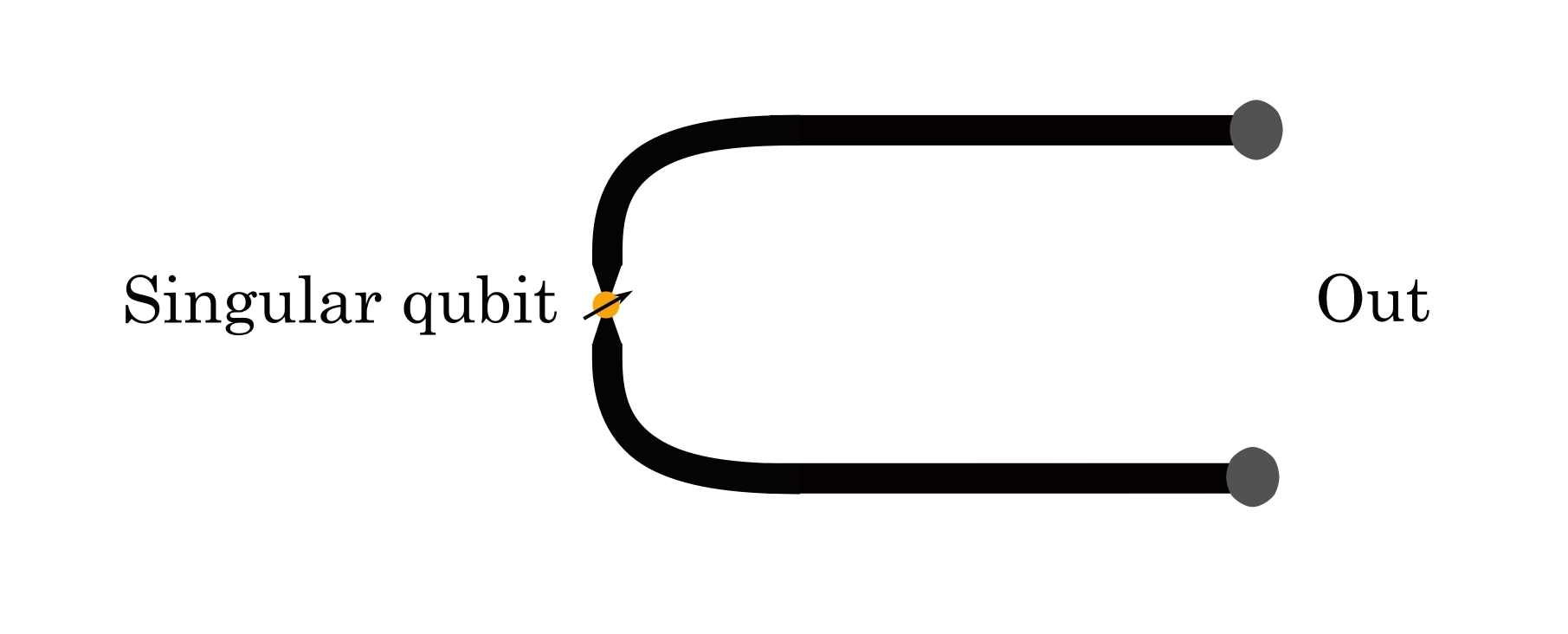}
	\caption{State injection in the Cluster state. The black lines represent the qubits that are measured in Z, they result in an encoded qubit on the output Toric code. The orange point is a punctual qubit that will be measured in $\frac{X+Y}{\sqrt{2}}$ or $Y$ basis, to enforce the $\ket{A}$ or $\ket{Y}$ state on the encoded qubit, respectively.}
	\label{fig:state injection in cluster}
\end{figure}

The logical state $\ket{A}$ cannot be created by Clifford gates and $X,Z$ creations and measurements. Figure~\ref{fig:state injection in cluster} illustrates the creation of a magic state: Here, the measurement lines must be shrunk on the initial plane to a single qubit, resulting in a Bell state between that qubit and the state on the `out' plane. A measurement of the single `in' qubit in the $\frac{X+Y}{\sqrt{2}}$ basis yields the desired encoded $\ket{A}$ state (or its orthogonal counterpart) on the `out' plane. Since the measurement lines must shrink, this procedure is not topologically protected and hence prone to errors. This will be resolved by subsequent distillations.

In~\cite{raussendorf_topological_2007}, the $S$ similarly required $\ket{Y}= \frac{\ket{0} + i \ket{1}}{\sqrt{2}}$ ancillas, see~(\ref{eq: S gate circuit}), which could only be obtained by the procedure above, but with a $Y$ measurement. The rebit encoding proposed here resolved that issue, except for the circuit~(\ref{A_RE}), which must be done with qubits. 

\begin{equation}
	\begin{quantikz}
		\lstick{$\ket{\psi}$} & \targ{} & \qw & \meter{Z} \\
		\lstick{$\ket{Y}$} & \control{} \vqw{-1} & \qw  & \qw &\rstick{$S\ket{\psi}$} \qw
	\end{quantikz}
	\label{eq: S gate circuit}
\end{equation}

In the following, we analyze the distillation circuit in details, and propose a circuit compactification that drastically improves the overhead of the $T$ gate.

\subsection{Magic state distillation}

To reduce the errors arising from the creation of the $\ket{A}$ state, one can resort to magic state distillation \cite{Bravi_Kitaev_magic_distillation}. In this paper, we focus on the methods using the 15 qubits Reed-Muller code \cite{steane1996quantumreedmullercodes, knill1996thresholdaccuracyquantumcomputation, Reed_Muller_website}. A Toric code state is first created fault-tolerantly by MBQC with 16 logical qubits in it. Those logical qubits are not created randomly but are eigenstates of very specific Pauli operators. The 15 first qubits are in the code space of the 15 qubit Reed-Muller code (they are eigenstates of 4 X stabilizers and 10 Z stabilizers, thus creating another level of encoding). This state behaves as an over encoded qubit which is, on top of that, entangled as a Bell pair with the 16th remaining qubit. This procedure involves only creation of individual qubit in the $\logicX$ and $\logicZ$ bases, and $\CNOT$ gates, and thus can be made fault-tolerant in MBQC. For the exact quantum circuit, we refer the reader to Ref.~\cite{raussendorf_topological_2007}, Fig. A.1. Translating it to MBQC, we show the corresponding measurement pattern in Fig. \ref{fig: circuit distillation} . The 16 pairs of lines ending on the right are the 16 logical qubits mentioned above. The overall state after this first step is 
\begin{equation}
    \ket{\varphi} = \frac{\ket{0^{RM}}\otimes\ket{0}_{16} + \ket{1^{RM}}\otimes\ket{1}_{16}}{\sqrt{2}}
\end{equation}
where $\ket{0^{RM}}$ and $\ket{1^{RM}}$ are eigenstates of the 15 qubits Reed-Muller code stabilizers and of the associated logical $\logicZ^{RM}$ operator for the value $+1$ and $-1$ respectively. 

Then, a transversal encoded $T$ gate is applied to the 15 qubits of the Reed-Muller code. This is done by injecting a magic $\ket{A}$ state like in Fig. \ref{fig:state injection in cluster} on each qubit pair of lines and performing the circuit in Eq. \ref{eq: T gate circuit annex}. The 15 Reed-Muller code is the smallest code where the $T^{RM}$ and $\logicX^{RM}$ gates are transversal. Therefore, the resulting state is still an eigenstate of the Reed-Muller stabilizers and entangled with the 16th qubit, but in a rotated basis: 

\begin{equation}
    T^{RM}\ket{\varphi} = \frac{\ket{0^{RM}}\otimes\ket{0}_{16} + e^{\frac{i\pi}{4}} \ket{1^{RM}}\otimes\ket{1}_{16}}{\sqrt{2}}
\end{equation}

Finally, the 15 first qubits are all measured individually in the $\logicX$ basis and thus the encoded state is measured in the $\logicX^{RM}$ basis too. That projects the 16th qubit on the $\ket{A}$ state or its complex conjugate depending on the measurement outcome, which can be fixed applying a $S$ gate. On top of it, the $\logicX$ measurement outcomes are used to infer the Reed-Muller X-type stabilizers. If an error is thereby detected, the state is discarded and the process is started again. Then, the kept $\ket{A}$ states have a reduced error probability compared to the previously injected ones. This procedure can be repeated many times and by concatenation, the error probability shrinks exponentially. 

Note that T gates are used to distillate the ancillas, therefore $S$ gates must be performed too (see Eq. \ref{eq: T gate circuit annex}). We will use $\ket{Y}$ state injection and Eq. \ref{eq: S gate circuit}. Thanks to the Reed-Muller stabilizers, the number of $S$ gates needed in total can be reduced from 15/2 to $1705/512 \sim 3.33$. Therefore, we need to inject, in average, that many $\ket{Y}$ state ancillas. Those ancillas are also noisy, so we need to distillate them too. The procedure is similar to what we just described for the $\ket{A}$ state, except that we use the 7 qubits Steane code instead, which is much less volume consuming.

\subsection{Overhead}

\subsubsection{Magic state distillation, error probability and overhead}

The distillation circuit fails if either one of the $T$ gates fails. In other words if one ancilla $\ket{A}$ from the previous round of distillation is faulty, or if the correction in the topological circuit fails. By topological circuit, we mean all the parts in the distillation procedure that can be implemented fault-tolerantly by MBQC in the cluster and on which we can do error correction. Then, the probability of failure of the round $l$ of distillation is : $15 \epsilon_{l-1}^A + \epsilon_{top}(L^A_{topo}, \lambda_{l-1}, d_{l-1})$, with $\epsilon_{l-1}^A$ the error probability of the ancillas at the previous round and $\epsilon_{top}(L^A_{topo}, \lambda_{l-1}, d_{l-1})$ the failure rate of the correction in the cluster. $L^A_{topo}$ is the minimal cluster size and $\lambda_{l-1}, d_{l-1}$ the scaling parameters to optimize. As we measure the Reed-Muller code stabilizers, we can detect some errors induced by faulty ancillas and we discard those cases. The remaining cases, where we don't know if an error has happened or not, have an error probability $\epsilon_{l}^A$ given in Eq. \ref{eq: recursion relation annex } at first order. The power 3 in $\epsilon_{l-1}^A$ is due the Reed-Muller error detection. The overhead for the $\ket{A}$ distillation at round $l$ can now be computed, involving the failure rate of the process and its volume (volume of the topological circuit + volume of the ancillas $\ket{A}$ and $\ket{Y}$ distilled in the previous round).

\begin{equation}
	\begin{split}
 \epsilon_{l}^A&= 35( \epsilon_{l-1}^A)^3 + \epsilon_{topo}(L_{Y}, \lambda_{l-1}, d_{l-1})\\
		O^A_l&= \frac{1}{1- 15 \epsilon_{l-1}^A - \epsilon_{topo}(L_{A}, \lambda_{l-1}, d_{l-1})} (15O^A_{l-1} + \frac{1705}{512}O^Y_{l-1} +  24V_{A} \lambda_{l-1}^3)
	\end{split}
	\label{eq: recursion relation annex }
\end{equation}

In the same fashion, we can derive the recursion relations for the $\ket{Y}$ state distillation (see Eq. \ref{eq: recursion relation Y distillation}). The error probability of the ancilla after the $l^{th}$ round of distillation depends on the performance of the Steane code correction, the error probability of the previous round and the failure rate of the topological circuit. The overhead involves the probability of failure of the distillation times the volume it takes. $\lambda_{l-1}$ and $d_{l-1}$ are taken to be the same as in Eq. \ref{eq: recursion relation annex } for the $\ket{A}$ state.

\begin{equation}
	\begin{split}
 \epsilon_{l}^Y&= 7( \epsilon_{l-1}^Y)^3 + \epsilon_{top}(L_{Y}, \lambda_{l-1}, d_{l-1})\\
		O^Y_l&= \frac{1}{1- 7 \epsilon_{l-1}^Y - \epsilon_{top}(L_{Y}, \lambda_{l-1}, d_{l-1})} (7O^Y_{l-1} +  24 V_{Y} \lambda_{l-1}^3)
	\end{split}
	\label{eq: recursion relation Y distillation}
\end{equation}

\subsubsection{T gate overhead}

Finally, we can compute the T gate overhead (see Eq. \ref{eq: overhead T gate magic Y appendix}). As the circuit involves magic states injection, we need to take into account the volume induced by the distillation and the ancillas error probabilities. After the round of distillation $l_{max}$ for both $\ket{Y}$ and $\ket{A}$ states, the volume taken by the T gate is given by the volume of the topological circuit plus the overheads for the ancillas (the 0.5 factor in front of $O^Y_{l_{max}}$ accounts for the S gate to be performed only one time over two in average). Then the probability of failure of the circuit is due to faulty ancillas or topological circuit. 
\begin{equation}
	O_T(\Omega)= \Omega(O^A_{l_{max}} +0.5O^Y_{l_{max}} + 36 (\lambda_{l_{max}})^3V_T)\text{exp}((\epsilon_{l_{max}}^A + \epsilon_{l_{max}}^Y + \epsilon_{top}(L_T, \lambda_{l_{max}}, d_{l_{max}}))\Omega)
	\label{eq: overhead T gate magic Y appendix}
\end{equation}

To find the optimal parameters, we need to minimize the overhead with respect to $l_{max}$ and all the rescaling parameters used in each round of distillation : $\{ \lambda_{l_{max}}, d_{l_{max}}, ..., \lambda_{0}, d_{0} \}$. 

In the rebit encoding, the T gate takes an additional $\ket{A}^*$ distilled state and more volume for the topological circuit. The overheads is then as follows, and needs to be optimized as well.
\begin{equation}
	O_T(\Omega)= \Omega(2O^A_{l_{max}} +1.5O^Y_{l_{max}} + 24 (\lambda_{l_{max}})^3V_{T, RE})\text{exp}((2\epsilon_{l_{max}}^A + 1.5\epsilon_{l_{max}}^Y + \epsilon_{top}(L_{T,RE}, \lambda_{l_{max}}, d_{l_{max}}))\Omega)
	\label{eq: overhead T gate magic Y RE appendix}
\end{equation}

Note that the $\ket{A}$ state is converted in the rebit encoding only after the last round of distillation. One could do it directly at the first round but that would require $2^l15^l$ ancillas instead of only $2\times 15^l$, and the topological circuit would be wider as due to the additional qubits necessary to the rebit encoding.

\subsubsection{Topological circuit for the A state distillation}

\begin{figure}[p]
	\centering
\begin{subfigure}{0.35\textwidth}
    \includegraphics[width=1\linewidth]{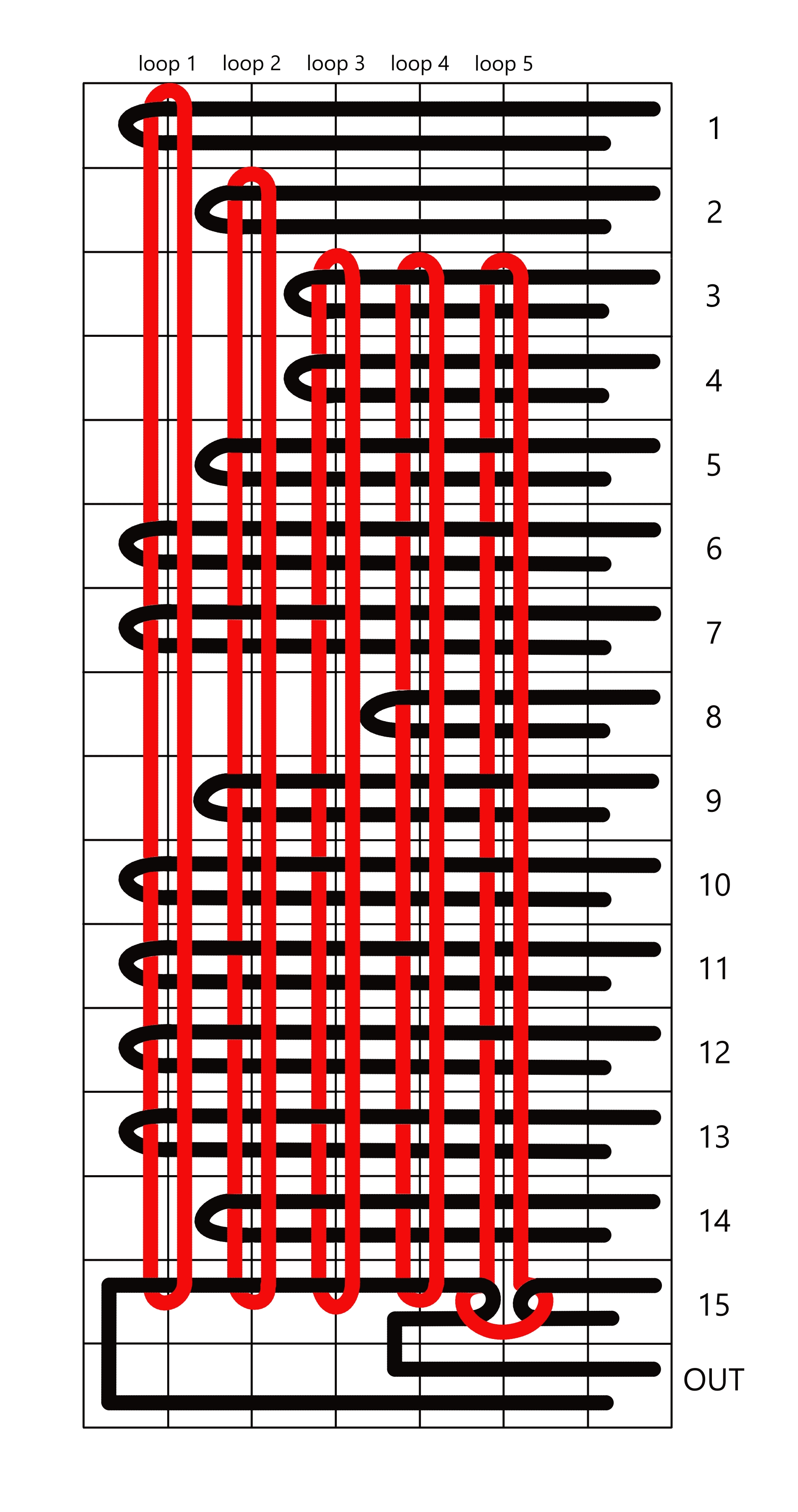}
    \caption{Topological circuit from Ref. \cite{raussendorf_topological_2007}. }
\end{subfigure}
\quad
\begin{subfigure}{0.6\textwidth}
    \includegraphics[width=1\linewidth]{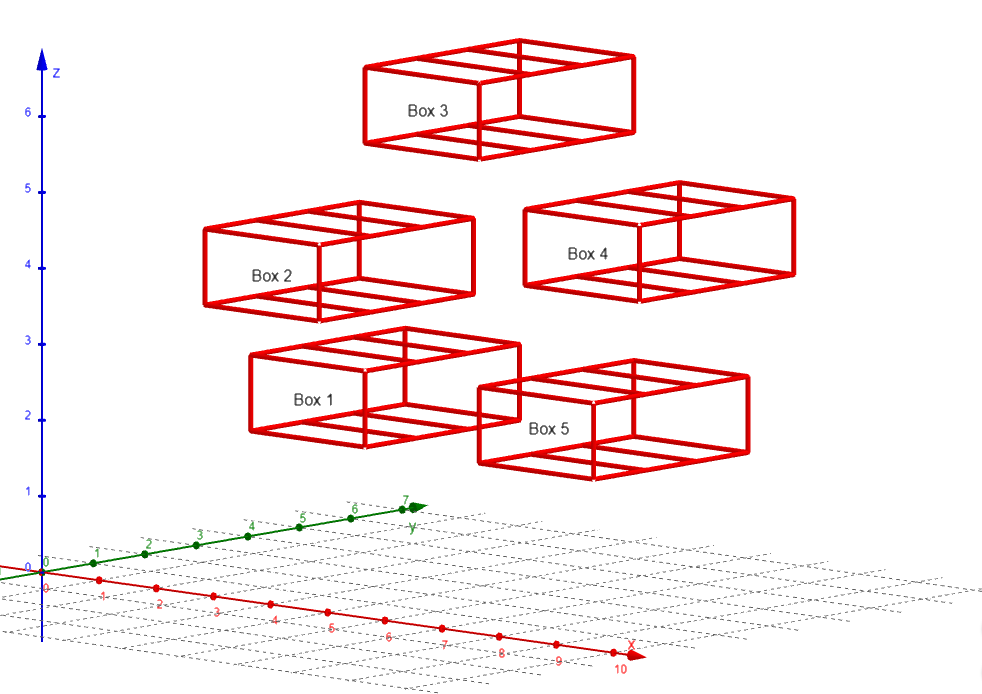}
    \caption{Optimized topological circuit. }
\end{subfigure}
\caption{Topological circuit of the $\ket{A}$ state distillation with the 15 qubits Reed-Muller encoding. (a) Reference circuit from Ref. \cite{raussendorf_topological_2007}. The grid in the background represents the unit cells of the lattice. (b) Circuit of the $\ket{A}$ sate distillation with the rectangular boxes introduce in Figure~\ref{fig: 8 lines in a rectangle cube}. There are five boxes corresponding to the five loops in (a). The black lines connecting the boxes and encoding the logical qubits are not represented here.}
    \label{fig: circuit distillation}
\end{figure}

The topological circuit for the magic $\ket{A}$ state distillation proposed in~\cite{raussendorf_topological_2007} is shown in Figure~\ref{fig: circuit distillation}(a). It is made only of qubit creations, $\CNOT$ gates which can be implemented in a topological fashion, see Figure~\ref{fig: CNot}. On the output side, the Reed-Muller state of the first 15 qubits are in a Bell-pair with the 16th qubit (`OUT' label in the figure): $\frac{1}{\sqrt{2}}\left( \ket{0^{RM}}\times\ket{0}_{\mathrm{out}} + \ket{1^{RM}}\times\ket{1}_{\mathrm{out}}\right)$.

The almost planar circuit of Figure~\ref{fig: circuit distillation}(a) requires a large volume. In order to improve this, we first use the equivalence introduced in Figure~\ref{fig:Big Equivalence}, then contract the loops into boxes as in~\ref{fig: 8 lines in a rectangle cube}, and finally optimize the relative positioning of five blocks. The result presented in Figure~\ref{fig: circuit distillation}(b) (showing all the red boxes but not the connecting black lines for clarity) spans 192 unit cells. The architecture mirrors the geometrical description of the 15 qubit Reed-Muller code in~\cite{Reed_Muller_website}. Indeed, the boxes implement the X stabilizers and they should be close to each other whenever they share a logical qubit, yielding a natural `tetrahedral' geometry. Both circuits presented in Fig. \ref{fig: circuit distillation} have been positively checked by our algorithmic verifier. We leave it to subsequent work to develop a computer code to search of the optimal architecture for any given circuit.

\newpage
\section{Algorithmic Circuit Verification}
\subsection{Mathematical foundation}

\subsubsection{Homology and relative homology} 

Recall that by $G_0, G_1, G_2,$ and $G_3$ we respectively denote the set of the lattice's primal vertices, edges, squares, and volumes, as well as that $C_n$ denotes the vector space associated with~$G_n$. $C_n$~is composed of formal linear combination of $n$-cells, i.e., of vectors $c = \sum_{g \in G_n} \, \alpha_g \, g$ with $\alpha_g \in \F_2$. We have then $C_n \cong \F_2^{m_n}$ with $m_n = |G_n|$, and the elements of $G_n$ can be represented by $\{e_i\}$, the canonical basis of~$\F_2^{m_n}$.

Furthermore, recall that the linear operators $\partial_n : C_n \rightarrow C_{n-1}$, called the \textit{boundary operators}, in particular satisfy~$\partial_n \circ \partial_{n-1} = 0$. With respect to the canonical basis described above, the boundary operators are $(m_{n-1}\times m_{n})$-matrices with $\{0,1\}$ entries, where $(\partial_n)_{ij} = 1$ if and only if the $i$th edge belongs to the boundary of the $j$th square.

The sequence $(C_n, \partial_n)$ forms a chain complex. Its dual chain complex~$(\overline{C}_n, \overline{\partial}_n)$ consists of its algebraic dual spaces and its dual maps, i.e., $\overline{C}_n = C_{\smash{3 - n}}^*$ and $\overline{\partial}_n = \partial_{\smash{3 - n}}^*$. Since all considered vector spaces are finite, they are naturally isomorphic to their respective dual counterparts, resulting in $\overline{C}_n \cong C_{3-n}$ (see Eq.~\ref{dual chain space}).

We now turn to the impact of qubit measurement lines, which correspond to a chain $l \in C_1$ and a cochain $\overline{l} \in \overline{C}_1 \cong C_2$, and effectively remove the edges and faces that have been measured from the primal, respectively dual chain complexes. This is best described in terms of relative homology, see~\cite{raussendorf_fault-tolerant_2006}, since relative cycles are chains of the original space whose boundaries are chains in the subspace.

For any chain $c = \sum_{i} \, \alpha_i e_i\in C_n$, we let 
\begin{equation}
    V_n(c) = \text{span} \{ e_i \in C_n : \alpha_i = 1\}
\end{equation}
be the $n$-chains made up of the measured qubits, and let $W_n(c) = C_n /  V_n(c)$ be the quotient space of $C_n$ by $V_n(c)$. $W_n(c)$ is naturally isomorphic to $ \text{span} \{ e_i \in C_n : \alpha_i = 0\}$ and we shall use this identification to write $C_n = V_n(c)\oplus W_n(c)$ and introduce the projection $P_n: C_n\to W_n(c)$ (equivalently: the quotient map). The original complex projects down to 
\begin{equation}
    C_3 \; \overset{\partial_3^r}{\longrightarrow} \; W_2(\bar l) \; \overset{\partial_2^r}{\longrightarrow} \; W_1(l) \; \overset{\partial_1^r}{\longrightarrow} \; C_0,
\end{equation}
where the relative boundary operator $\partial^r_n$ is given by $\partial^r_n = P_{n-1} \partial_n P_n$. 

In this new complex, two chains are equivalent if they differ by a relative boundary, namely $c-c' \in \mathrm{Ran} (\partial^r)$. Concretely in terms of the original complex and in the case $n=1$, two lines are equivalent if their difference is a sum of the boundary of a surface and a measured line (since the latter is in the kernel of $\partial^r$).

In terms of the matrix representatives of the boundary maps, $\partial^r_n$ is obtained from $\partial_n$ by setting those rows and columns to zero whenever they correspond to measured qubits. Equivalently by simply deleting these rows and columns, which amounts to considering $\partial^r_n: \mathrm{Ran} (P_n)\to\mathrm{Ran} (P_{n-1})$.

\subsubsection{Correlation surfaces} 

The algorithm's design is limited to CSS gates, namely those unitaries in Theorem~\ref{Thm: Fundamental} for which (\ref{eq: X condition on OUT surface}) and (\ref{eq: Z condition on OUT surface}) hold without the $Z$, respectively $X$ operator on their left hand side. Accordingly, the equations for primal and dual surfaces are separated and (\ref{eq: X condition on IN surface}, \ref{eq: X condition on OUT surface}) can be treated independently of (\ref{eq: Z condition on IN surface}, \ref{eq: Z condition on OUT surface}). 

Verifying that a measurement pattern specified by $l$ corresponds to a circuit represented by the primal surface boundaries $c_1, \dots, c_{n} \in C_1$ on the `in' plane $\mathcal{I}$ and the `out' plane $\mathcal{O}$ means checking whether correlation surfaces with these boundaries exist. In terms of Theorem~\ref{Thm: Fundamental}, this means finding solutions $c_i^Z$ of~(\ref{eq: Z condition on IN surface},\ref{eq: Z condition on OUT surface}) such that $\partial_2^r c_i^Z \cap (\mathcal{I}\cup\mathcal{O})= c_i$. The use of the relative boundary operator in this equation ensures that the solution surfaces are supported by the measurement lines inside the cluster. This is equivalent to verifying whether all of the linear systems of equations
\begin{equation}
    \partial_2^r x = c_i
\end{equation}
for $i=1,\ldots,n$ have a solution. The Rouch\'e-Capelli theorem ensures that this holds if the matrix $\partial_2^r$ has the same rank as the augmented matrices $( \partial_2^r \, | \, c_{i})$. The same must hold for dual surfaces and their boundaries, yielding the additional equations 
\begin{equation}
    \overline{\partial}_2^r y = \overline{c}_{j}
\end{equation}
and the corresponding rank-computing problem for $\overline \partial_2^r$ and $(\overline \partial_2^r \, | \, \overline c_{i})$.

Summarizing, after the preliminary steps of building the lattice, constructing the boundary and relative boundary matrices described above, the verification algorithm reduces to the standard linear algebra task of computing the rank (over $\mathbb{F}_2$) of $2(n+1)$ matrices. The whole process, which we describe below in detail, can be done in polynomial time in the linear size of the cluster.

\subsection{Algorithmic Verification}

We assume the lattice to be a rectangular cuboid whose side lengths (or shape) is given by $(s_1, s_2, s_3) \in \mathbb{N}^3$. The most decisive design choice is selecting how to represent the sets $G_n$ numerically. We place the $m_0 = \prod_i s_i$ primal vertices along the equidistant points $(x_1, x_2, x_3) \in [s_1] \times [s_2] \times [s_3]$. The $m_3 = \prod_i (s_i - 1)$ primal cubes can then be represented by their midpoints, i.e., by $(z_1, z_2, z_3) = (y_1 + 0.5, y_2 + 0.5, y_3 + 0.5)$ with $y_i \in [s_i - 1]$. The $m_1$ primal links and $m_2$ primal faces are subsequently characterized by the primal sites or primal cubes they respectively connect; specifically, by the midpoint of these primal sites or primal cubes. One advantage of these human readable representations is that they allow for manual data entry of circuits to verify. Foremost, they enable a straightforward computation of the (dual) boundary of an object. For example, the boundary of a primal link $(x_1, x_2 + 0.5, x_3) \in G_2$ is given by
\begin{align}
    \{ (x_1, x_2, x_3), (x_1, x_2 + 1, x_3) \} \subset G_1,
\end{align}
and the boundary of a primal face {$(z_1 + 0.5, z_2, z_3) \in G_3$} is given by
\begin{align}
\begin{aligned}
    \{
    & (z_1 + 0.5, \, z_2 - 0.5, \, z_3), (z_1 + 0.5, \, z_2, \, z_3 - 0.5), \\
    & (z_1 + 0.5, \, z_2 + 0.5, \, z_3), (z_1 + 0.5, \, z_2, \, z_3 + 0.5)
    \} \subset G_2.
\end{aligned}
\end{align}
This generalizes trivially to {links $(x_1 + 0.5, x_2, x_3), (x_1, x_2, x_3 + 0.5) \in G_2$} and faces $(z_1, z_2 + 0.5, z_3)$, $(z_1, z_2, z_3 + 0.5) \in G_3$.

These preliminaries enable us to provide a high-level outline of the algorithm. We assume to be given the shape $(s_1, s_2, s_3)$ of the lattice, along with the sets of qubit measurement lines~$L \subset G_1$ and~$\overline{L} \subset \overline{G}_1 = G_2$. Lastly, two lists consisting of primal surface boundaries $B_1, \dots, B_{n_1} \subset G_1$ and dual surface boundaries $\overline{B}_1, \dots, \overline{B}_{n_2} \subset \overline{G}_1 = G_2$ to verify are required. The program then proceeds as follows.

\begin{enumerate}
    \item \textbf{Building the lattice:} For convenience, the representations of all $n$-cells are stored in an iterable data type. Afterwards, a bijection between the sets $G_n$ and their respective index sets $[m_n]$ can be created by simply enumerating all objects in $G_n$ and storing the result in a dictionary. All of this is done once in time $\mathcal{O}\big( \text{poly}(s) \big)$ where $s = \|(s_1, s_2, s_3)\|_\infty$. Since $s$ is typically of order~$< \! 10^2$, optimizing polynomial run-times is not a dominant concern, both here and in the following. Further note that accessing elements in such an associative array can be done in time $\mathcal{O}(1)$, i.e., switching between an $n$-cell's descriptive representation in $G_n$ and its index in $[m_n]$ or vector in $\F_2^{m_n}$ in subsequent calculations introduces a negligible computational overhead at all reasonable problem sizes.

    \item \textbf{Computing the relevant boundary operators:} The matrix representation of an operator $\partial_n$ can be computed by calculating its action on all basis vectors $e_i$ of $C_n \cong \F_2^{m_n}$. Said action can be determined by mapping $e_i$ to its representation in $G_n$, computing the boundary with that representation (as exemplified above), and mapping the result back from $G_{n - 1}$ to $C_{n - 1} \cong \F_2^{m_{n - 1}}$. In our concrete case, where we are ultimately interested in the operators $\partial^r_2$ and $\overline{\partial}^{\smash{r}}_2$, this is done for $n \in \{1, 2\}$, as the matrix representation of $\overline{\partial}_2$ is given by the transpose of the matrix representation of $\partial_1$. This step hence generates matrices $\partial_1$ and $\partial_2$ with $(\partial_n)_{ij} = 1$ if and only if the $i$-th $(n-1)$-cell is in the boundary of the $j$-th $n$-cell, again requiring time~$\mathcal{O}\big( \text{poly}(s) \big)$.

    \item \textbf{Computing the respective \textit{relative} boundary operators:} Given a boundary operator $\partial_n$, the corresponding relative boundary operator $\partial^r_n$ acting on $W_n \cong C_n/V_n$ is calculated with $\partial^r_n = P_{n-1} \, \partial_n \, P_n$. The projectors $P_n$ are dependent on the qubit measurement lines $L$ and $\overline{L}$, and can be calculated by removing the $i$-th diagonal element of an identity matrix if and only if the $i$-th $n$-cell is an element of $L \cup \overline{L}$. Alternatively, to avoid storing the projectors in memory, one can iterate through all elements in $L \cup \overline{L}$ and equate to zero the possibly corresponding row or column of $\partial_n$. This approach is taken by us, again for $n \in \{1, 2\}$ while requiring time~$\mathcal{O}\big( \text{poly}(s) \big)$. Note that the dual lattice must also be `smooth', there cannot be half dual unit cells on the cluster boundary, which can effectively be implemented and ignoring the outer-most primal links by removing relevant entries from the dual relative boundary operator.

    \item \textbf{Computing the ranks of the (augmented) relative boundary operators:} After using the created dictionary to map the surface boundaries $B_1, \dots, B_{n_1} \subset G_1$ and $\overline{B}_1, \dots, \overline{B}_{n_2} \subset \overline{G}_1 = G_2$ to the target vectors $c_1, \dots, c_{n_1} \in C_1$ and $\overline{c}_1, \dots, \overline{c}_{n_2} \in \overline{C}_1$, the ranks of $\partial_2^r,\allowbreak ( \partial_2^r \, | \, c_1),\allowbreak \dots,\allowbreak ( \partial_2^r \, | \, c_{n_1}),\allowbreak \bar{\partial}_2^r,\allowbreak (\bar{\partial}_2^r \, | \, \bar{c}_1),\allowbreak \dots,\allowbreak ( \bar{\partial}_2^r \, | \, \bar{c}_{n_2})$ can be calculated in time~$\mathcal{O}\big( \text{poly}(s) \big)$ by, for example, using any linear algebra library operating over finite fields. We employed the \textit{galois} package \cite{Hostetter_Galois_2020} for Python.

\end{enumerate}

\end{document}